\documentclass[aps,prl,reprint,amsmath,amssymb,superscriptaddress,nofootinbib,longbibliography]{revtex4-2}
\usepackage{graphicx}
\usepackage[colorlinks=true,urlcolor=blue,citecolor=blue,linkcolor=blue,bookmarks=false, pdfstartview={FitH}]{hyperref}
\usepackage{amsmath}
\usepackage{bm}
\usepackage{braket}
\usepackage{dcolumn}
\usepackage{textcomp}
\usepackage{gensymb}
\usepackage{physics}
\usepackage{color,soul}
\usepackage{ragged2e}
\usepackage{color,soul}
\def\BiSe{Bi$_2$Se$_3$\,}
\def\BiTeSe{Bi$_2$Te$_2$Se\,}
\def\BiTe{Bi$_2$Te$_3$\,}
\def\BiAll{Bi$_2$Te$_{3-x}$Se$_{x}$\,}

\begin{document}

\title{Electronic and vibrational excitations on the surface of the three-dimensional topological insulator \BiAll ($x = 0, 2, 3$)}
\author{A.~Lee}
\email{aclee@physics.rutgers.edu}
\affiliation{Department of Physics \& Astronomy, Rutgers University, Piscataway, NJ 08854, USA}
\author{H.-H.~Kung}
\altaffiliation[Current affiliation: ]{Quantum Matter Institute, University of British Columbia, Vancouver, BC V6T 1Z4, Canada}
\affiliation{Department of Physics \& Astronomy, Rutgers University, Piscataway, NJ 08854, USA}
\author{Xueyun Wang}
\affiliation{Department of Physics \& Astronomy, Rutgers University, Piscataway, NJ 08854, USA}
\affiliation{Rutgers Center for Emergent Materials, Rutgers University, Piscataway, NJ 08854, USA}
\author{S.-W.~Cheong}
\affiliation{Department of Physics \& Astronomy, Rutgers University, Piscataway, NJ 08854, USA}
\affiliation{Rutgers Center for Emergent Materials, Rutgers University, Piscataway, NJ 08854, USA}
\author{G.~Blumberg}
\email{girsh@physics.rutgers.edu}
\affiliation{Department of Physics \& Astronomy, Rutgers University, Piscataway, NJ 08854, USA}
\affiliation{National Institute of Chemical Physics and Biophysics, 12618 Tallinn, Estonia}


\begin{abstract}
We study surface states in the three-dimensional topological insulators \BiAll ($x = 0, 2, 3$) by polarization resolved resonant Raman spectroscopy. 
By tracking the spectral intensity of the surface phonon modes with respect to the incident photon energy, we show that the surface phonons are qualitatively similar to their bulk counterparts. 
Using the resonant Raman excitation profile, we estimated the energy gaps between the surface conduction bands and the bulk conduction bands. 
In addition, we selectively excite the surface-to-bulk electronic continuum near the Fermi energy in \BiSe to determine the strength of Fano interaction between the  most prominent surface phonon and the surface-to-bulk continuum. 
\end{abstract}

\maketitle

\section{Introduction}
Three-dimensional topological insulators (3DTIs) are materials that contain gapless, Dirac-like electronic surface states~\cite{moore2010birth,hasan2010colloquium,fu2007topological} that are topologically protected against elastic scattering by non-magnetic defects~\cite{roushan2009topological} making them attractive candidates to the application of devices where coherent spin-polarized transport is desired~\cite{zuti2004spintronics}. 
Given that the electronic surface states in 3DTIs are confined to the topmost unit cells~\cite{linder2009anomalous}, complete understanding of the lattice dynamics interacting with the surface states is critical. 
This knowledge is essential not only because the abrupt surface termination in materials can radically alter the crystal structure and electronic properties at the surface, but also because inelastic scattering from surface to bulk electronic states via phonons is expected to be the dominant low-energy scattering mechanism at room temperature~\cite{park2010quasiparticle}. 
For these reasons, we investigated the surface phonons in the 3DTIs \BiAll ($x = 0, 2, 3$) using polarization-resolved resonant Raman spectroscopy to elucidate their interaction with the electronic states. 

The bismuth-based 3DTIs \BiAll are extensively studied topological insulator systems due to their stoichiometry, large bulk band gap, and simple surface spectrum. 
\BiAll are composed of quintuple layers (QL) of covalently bonded Bi, Te, and Se atoms stacked along the c-axis~\cite{nakajima1963crystal} with $D_{3d}$ (R$\bar{3}$m) point group symmetry~\cite{nakajima1963crystal}. 
The symmetry of the topmost QL is reduced from the 3D R$\bar{3}$m group to the two-dimensional (2D) $C_{6v}$ (P6mm) wallpaper group. 
The QL are weakly bound together by van der Waals forces resulting in a quasi-2d electronic band structure~\cite{ribak2016internal,arakane2012tunable,chen2009experimental}. 
The principal bulk band gap is found at the $\Gamma$-point. 
The lowest bulk conduction band is denoted as CB0. 
The highest occupied Dirac-cone, $SS1$, is composed of electronic surface states, and lies within the principal bulk band gap, where the bands cross at the $\bar{\Gamma}$-point~\cite{xia2009observation,arakane2012tunable,chen2009experimental}. 
There is also an unoccupied Dirac cone that lies between the second- and third-lowest bulk conduction bands~\cite{niesner2012unoccupied}. 

In this work, we apply the polarization-resolved resonant Raman excitation profile (RREP) spectroscopy of the bulk and surface phonon modes in the bismuth based topological insulators \BiAll to study excitations between the bulk and surface bands. 
The surface phonons were identified and characterized by using polarization-resolved measurements at the incident photon energies that maximize the surface phonon intensity. 
We detect the relative energy gaps between the surface bands by analyzing the intensity of the phonon modes as a function of incident photon energy.  
We selectively excite the lowest energy electron-hole excitation continuum due to transitions from surface to bulk electronic states that interacts with the surface phonons. 
For in \BiSe, by analyzing the interference Fano lineshape of the surface phonons with the continuum as a function of excitation energy, we establish the electron-phonon coupling strength between the surface phonons and the electronic surface states near the Fermi energy. 
This work also extends the earlier studies of surface phonons in \BiSe~\cite{kung2017surface} and \BiTe~\cite{boulares2018surface} as it includes resonant Raman enhancement effects, as well as results for \BiTeSe. 

\section{Experimental Details}
Single crystals of \BiAll were grown by the modified Bridgman technique~\cite{dai2016toward}. 
Although the chemical potential in ideal topological insulators rests at the intersection point of the Dirac cones within the principal bulk band gap, the presence of natural defects typically raises the chemical potential into the upper Dirac cone and CB0~\cite{dai2016toward,wang2013native}, thus,  
the lowest conduction bands for all samples are assumed to be occupied. 
The chemical potential of the \BiSe samples grown for this study were determined by scanning tunneling spectroscopy to cross approximately 150\,meV above the Dirac point of the principal surface band~\cite{dai2016toward} and was later confirmed also by the optical studies~\cite{kung2017chiral,kung2019observation}. 
All samples were cleaved in a N$_2$-rich glove bag and then in-situ transferred to a continuous He-flow optical cryostat. 

\begin{table}[t]
	\caption{\label{table:irreps}The Raman selection rules in the bulk and on the surface of \BiAll ($x = 0, 2, 3$).}
	\setlength{\tabcolsep}{8pt}
	\setlength{\extrarowheight}{3pt}
	\vspace{5pt}
	\begin{tabular}{l l l}
		\hline \hline
		Scattering & Bulk & Surface \\ 
		Geometry & ($D_{3d}$) &  (C$_{6v}$) \\
		\hline
		XX & $A_{1g}$ + $E_{g}$ & $A_{1}$ + $E$\\
		XY & $A_{2g}$ + $E_{g}$ & $A_{2}$ + $E$\\ 
		RR & $A_{1g}$ + $A_{2g}$ & $A_{1}$ + $A_{2}$\\
		RL & $2E_{g}$ & $2E$ \\
		\hline \hline
	\end{tabular} 
\end{table}

\begin{figure}[t]%
	\centering	\includegraphics[width=\linewidth]{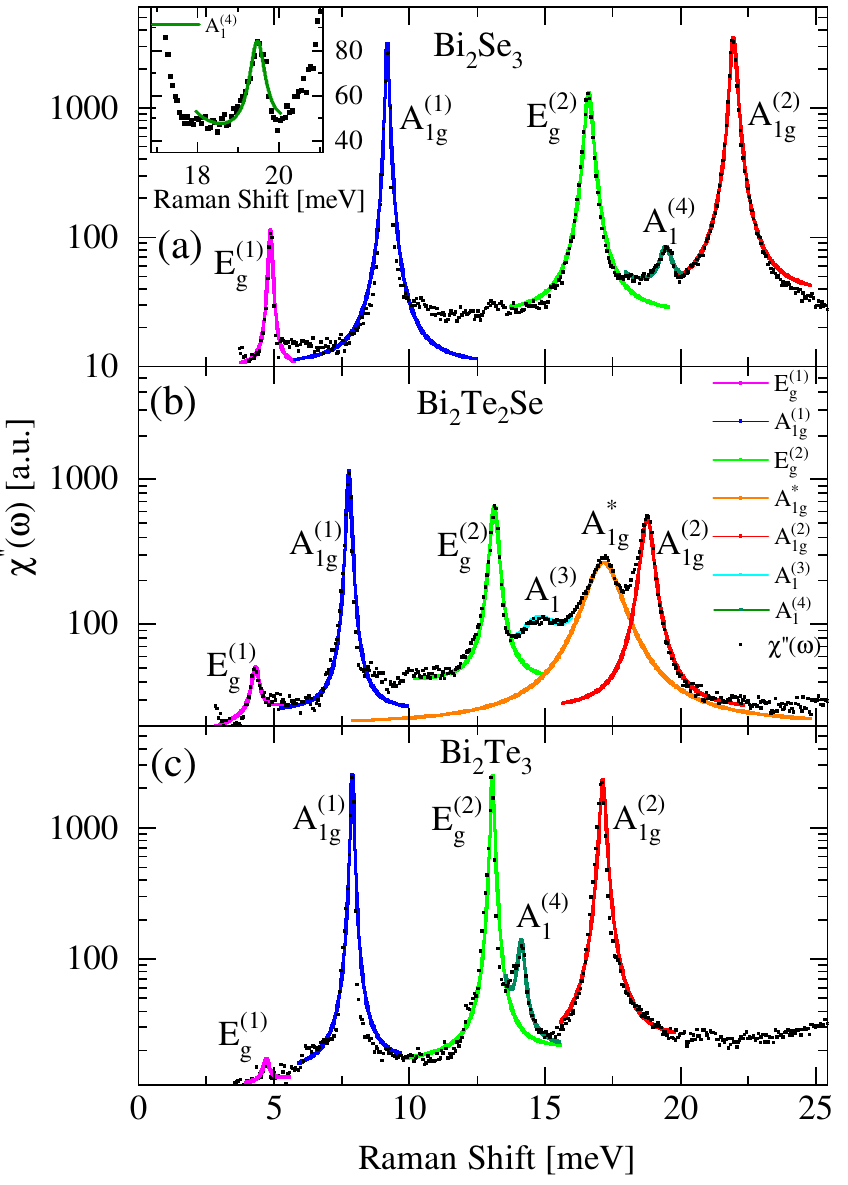}
	\caption{Raman response spectra of \BiAll in the XX scattering polarization geometry taken from (a) \BiSe, (b) \BiTeSe, and (c) \BiTe.
		(inset) Data from \BiSe plotted against linear scale to illustrate signal-to-noise ratio. 
		All measurements were performed at 15\,K with $\omega_{i} =$ 1.92\,eV excitation. 
		The bulk phonons are overlaid with a fit to Lorentzian lineshape over a local linear background for visual aid. 
		The mode labeled A$_{1g}^{*}$ is an antisite defect mode. 
        The secondary emission spectra were normalized for wavelength dependent scattering volume, the attenuation in the sample, and reflectivity/transmission of light at the surface interface, see Eqs.\,(\ref{SE}) and (\ref{Corr}), using optical data from Refs.\,\cite{mciver2012theoretical,aleshchenko2014infrared,cui1999optical}; the Raman susceptibility was derived by subtracting the PL signal. 
	}
\label{overview}
\end{figure}

Polarization-resolved Raman scattering measurements were performed in the quasi-backscattering geometry. 
For excitation, we use Kr$^{+}$ ion laser lines with photon energies ranging from 1.55 to 3.05\,eV. 
All spectra were acquired with 1800\,grooves$/$mm diffraction gratings (spectral resolution $\sim0.1$\,meV) using a triple stage spectrometer setup.  

The scattering polarization geometries used in this paper are labeled as $e_{i}e_{s}$, where $e_{i(s)}$ denotes the polarization of the incident (collected) light. 
The incident and collected photons propagate along the c-axis, with the polarization directed in-plane. 
The energy of the incident (scattered) photons are denoted as $\omega_{i(s)}$. 
The four scattering polarization geometries used in this paper are XX, YX, RR, and RL. 
The linear light polarization directions, X and Y, lie within the ab-plane and are orthogonal to one another. 
R and L denote right and left circularly polarized light, respectively, such that $R(L) = X \pm iY$. 

Raman selection rules are listed in Table\,\ref{table:irreps}.
The bulk Raman excitations have either $A_{1g}$, $A_{2g}$, or $E_{g}$ symmetry. 
The irreducible representations of the Raman-active phonons at the $\Gamma$ point are $2A_{1g} + 2E_{g}$~\cite{richter1977raman,wang2012phonon,cheng2011phonons}. 
In addition, there are four infrared (IR) active phonons, whose irreducible representations are $2A_{2u} + 2E_{u}$~\cite{richter1977raman}. 

\begin{table}[t]
	\caption{\label{table:mode} Summary of the bulk and surface mode energies in \BiTeSe and \BiTe. All values are given in units of meV.}
	\setlength{\tabcolsep}{0.75pt}
	\setlength{\extrarowheight}{3pt}
	\vspace{3pt}
	\begin{tabular}{l c c c c c}
		\hline\hline
		{} & \multicolumn{2}{c}{Bi$_2$Te$_2$Se} & { } & \multicolumn{2}{c}{Bi$_2$Te$_3$} \\
		\cline{2-3} \cline{5-6}
		Symmetry & \multicolumn{1}{c}{This work} & \multicolumn{1}{c}{Literature} & & \multicolumn{1}{c}{This work} & \multicolumn{1}{c}{Literature} \\
		\hline 
		$A_{1g}^{(1)}$ & 8.1 & 8.4~\cite{akrap2012optical} & & 7.8 & 7.8~\cite{richter1977raman} \\ 
		$A_{1g}^{(2)}$ & 19.1 & 19.6~\cite{akrap2012optical} & & 17.2 & 16.6~\cite{richter1977raman} \\
		$E_{g}^{(1)}$ & 4.5 & 4.3~\cite{akrap2012optical} & & 4.7 & 4.4~\cite{wang2012phonon} \\
		$E_{g}^{(2)}$ & 13.5 & 13.7~\cite{akrap2012optical} & & 13.1 & 12.8~\cite{richter1977raman} \\
		$A_{2u}^{(1)}$ & -- & 15.9~\cite{shi2015connecting} & & -- & 11.7~\cite{richter1977raman}\\
		$A_{2u}^{(2)}$ & -- & 17.1~\cite{shi2015connecting} & & -- & 14.9~\cite{richter1977raman}\\
		$E_{u}^{(1)}$ & -- & 7.7~\cite{reijnders2014optical} & & -- & 6.2~\cite{richter1977raman} 6.0~\cite{boulares2018surface,wang2012phonon}\\
		$E_{u}^{(2)}$ & -- & 14.5~\cite{akrap2012optical,dipietro2012optical} & & -- & 11.8~\cite{richter1977raman}\\
		\hline
		$A_{1}^{(1)}$ & 7.4 & -- & & 7.5 & 7.7~\cite{ren2012large} \\
		$A_{1}^{(2)}$ & -- & -- & & 16.7 & 17.4~\cite{ren2012large} \\
		$A_{1}^{(3)}$ & 15.2 & 14.8~\cite{tian2016local} & & 10.3 & -- \\
		$A_{1}^{(4)}$ & 16.7 & -- & & 14.0 & 14.1~\cite{boulares2018surface} 14.8~\cite{ren2012large} \\
		$E^{(1)}$ & 4.18 & -- & & -- & 5.0~\cite{ren2012large} \\
		$E^{(2)}$ & 13.0 & -- & & 12.3 & 12.4~\cite{ren2012large} \\
		$E^{(3)}$ & 7.8 & -- & & 7.1 & -- \\
		$E^{(4)}$ & 14.9 & -- & & 12.3 & -- \\
		\hline\hline
	\end{tabular}
\end{table}

\begin{figure}[t]%
	\centering	\includegraphics[width=\linewidth]{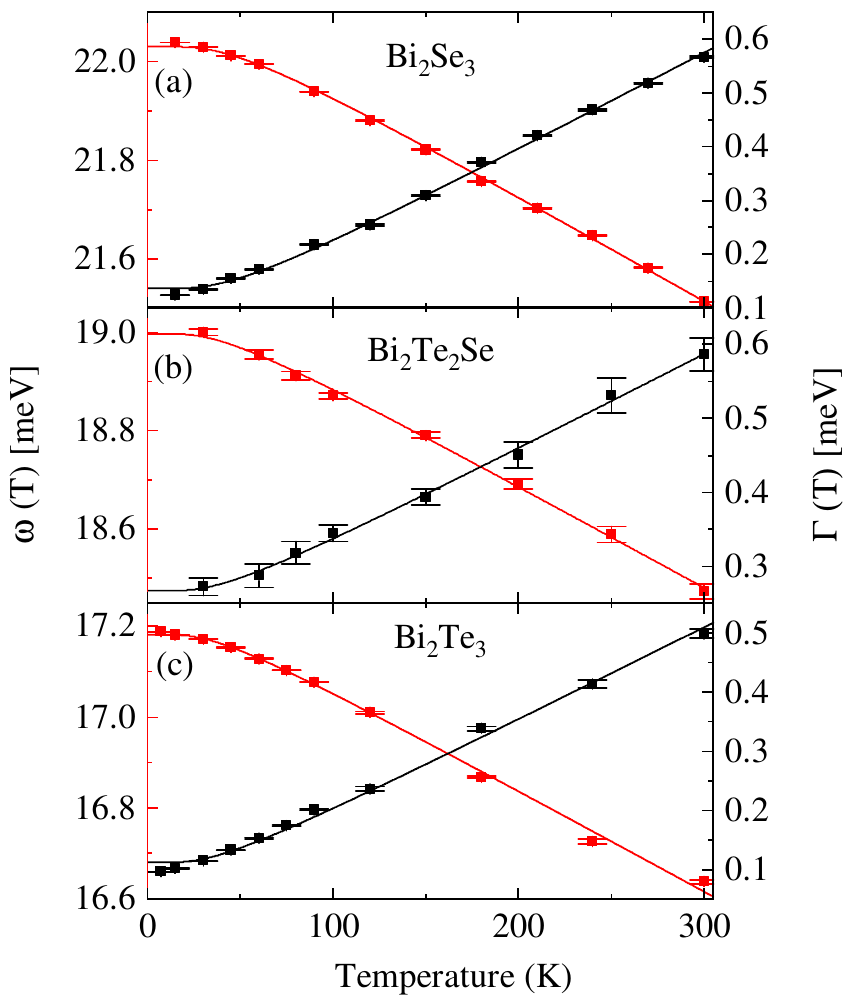}
	\caption{Temperature dependence of the energy and linewidth of the A$_{1g}^{(2)}$ phonon mode of \BiAll. 
	The energy (red) and linewidth (black) values of the A$_{1g}^{(2)}$ mode of (a) \BiSe, (b) \BiTeSe, and (c) \BiTe are displayed in units of meV.  
	The temperature dependence of the energy and linewidth of the A$_{1g}^{(2)}$ mode is fit to a 2-phonon anharmonic decay model~\cite{klemens1966anharmonic}. 
	}
\label{temperaturedep}
\end{figure}

In Fig.\,\ref{overview}, we plot on semi-log scale spectra of the Raman susceptibility for \BiAll. 
Measurements were performed in the XX scattering geometry at 15\,K. 
The major spectral features are fitted to a Lorentzian function and overlaid with the experimental data to aid the reader's eye. 

The secondary emission spectra contains both Raman and photoluminescence (PL) contributions. 
The measured secondary emission spectra, $S(\omega_i,\omega_s)$, differs from the Raman susceptibility, $\chi^{''}(\omega)$, by two factors
\begin{eqnarray}
\label{SE}
S(\omega_i,\omega_s) \sim C(\omega_i,\omega_s)[\{1 + n_B(\omega)\}\chi^{''}(\omega_i,\omega) \nonumber\\ + PL(\omega_i,\omega_{s})]
\end{eqnarray} 
where $n_{B}(\omega)$ is the Bose-Einstein distribution, $C(\omega_i,\omega_s)$ is a correction factor that accounts for the scattering volume and power loss due to the optical properties of the materials, and $PL(\omega_i,\omega_{s})$ is the contribution from photoluminescence. 
The form of $C(\omega_i,\omega_s)$ for the backscattering geometry was derived in Refs.\,\cite{loudon1965theory,blumberg1994investigation}
\begin{equation}
\label{Corr}
C(\omega_i,\omega_s) = \frac{\alpha(\omega_{i}) + \alpha(\omega_{s})}{T(\omega_{i})T(\omega_{s})}n(\omega_{s})^{2}
\end{equation} 
where $\alpha$ is the optical extinction coefficient, $T$ is the transmission coefficient at the crystal-vacuum interface, and $n(\omega_{s})$ is the index of refraction of the material. 
The values of the optical properties of the \BiAll crystals are taken from previously published data~\cite{mciver2012theoretical,aleshchenko2014infrared,cui1999optical}. 

The Raman susceptibility spectra was derived from the renormalized secondary emission spectra by subtracting the PL contribution. 
The PL contribution to the secondary emission spectra is approximately a lineary background in the narrow spectral range of the Raman excitations~\cite{kung2019observation}. 
We approximated the background to be the normalized spectral intensity 2.5\,meV away from the laser excitation line, where the Raman continuum is still negligibly weak.

\section{Results and Discussion}
\begin{figure*}[t]%
	\centering
	\includegraphics[width=0.9\linewidth]{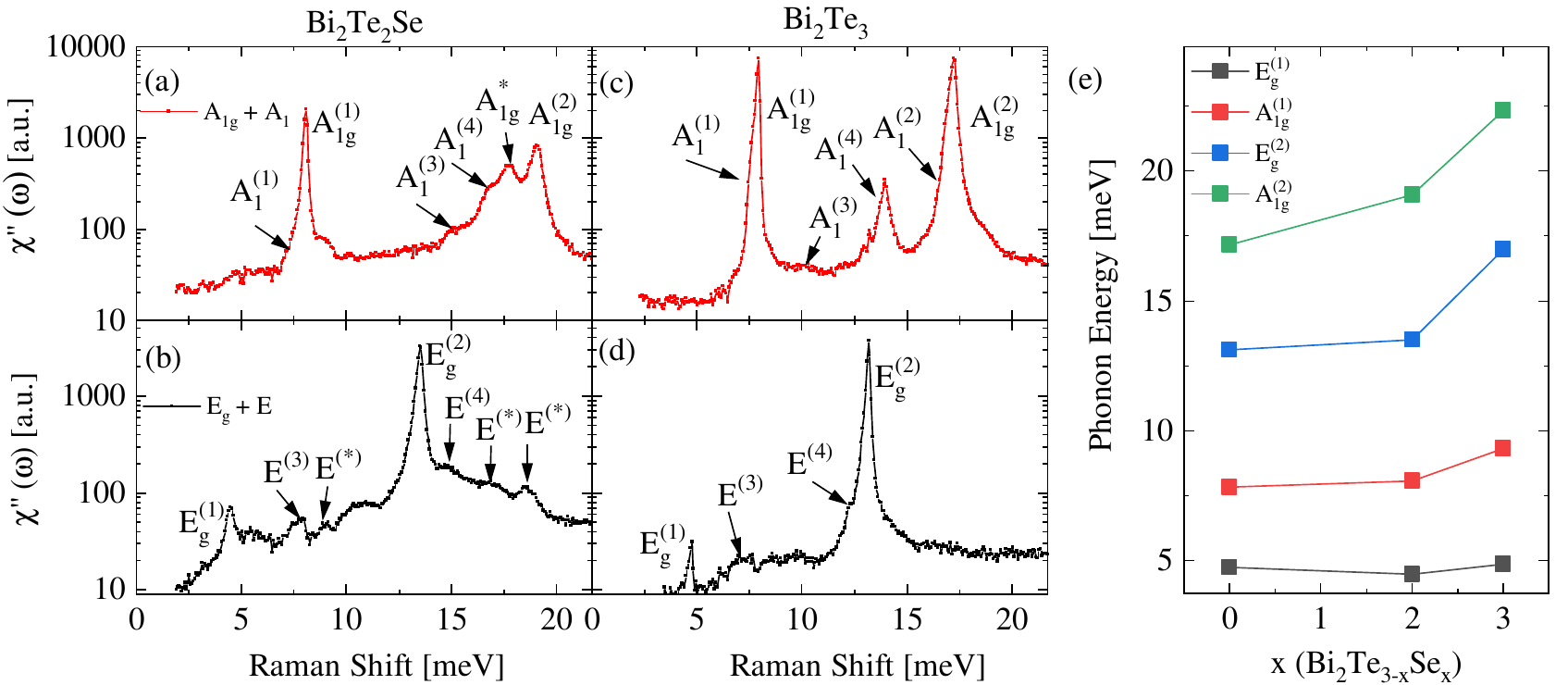}
	\caption{High resolution $\chi^{''}(\omega)$ spectra of \BiTeSe and \BiTe. 
		$\chi^{''}(\omega)$ spectra of \BiTeSe are taken in the (a) RR and (b) RL scattering geometries. 
		The modes labeled $E^{(*)}$ were previously reported~\cite{akrap2012optical}, though their origin is unclear. 
		$\chi^{''}(\omega)$ spectra of \BiTe are acquired in the (c) RR and (d) RL scattering geometries. 
		In all cases, the spectral were acquired at 15\,K with 1.92\,eV excitation energy, and are plotted in semi-log scale. 
		(e) Phonon energy of \BiAll as a function of Se concentration ($x$). 
	}
\label{surfacephonons}%
\end{figure*}

The most prominent spectral features in the \BiAll secondary emission spectra are the Raman-active bulk phonons. 
The energy of the modes are consistent with previously reported results, see Table\,\ref{table:mode}.
In \BiTeSe, there is an additional broad mode, labeled $A_{1g}^{*}$, that is generally accepted to be the result of an anti-site defect induced local vibration~\cite{tian2016local}. 
It should be noted that the bulk phonons consistently shifts to higher energies with Se concentration, see Fig.\,\ref{surfacephonons}(e). 

\begin{table}[b]
	\caption{\label{table:I}Residual linewidths $\Gamma_{0}$ for the $A_{1g}^{(2)}$ phonon modes. All values are given in units of meV.}
	\setlength{\tabcolsep}{8pt}
	\setlength{\extrarowheight}{3pt}
	\vspace{5pt}
	\begin{tabular}{l c c}
		\hline \hline
		Material & This work & Literature\\
		\hline
		\BiSe & 0.08 & 0.06~\cite{tian2017understanding}\\
		\BiTeSe & 0.20 & 0.39~\cite{tian2016local}\\
		\BiTe & 0.07 & 0.09~\cite{tian2017understanding}\\
		\hline \hline
	\end{tabular} 
\end{table}

In Fig.\,\ref{temperaturedep}, we show the evolution of the energy and linewidth of the $A_{1g}^{(2)}$ mode in \BiAll as a function of temperature. 
The $A_{1g}^{(2)}$ mode is used as an example of the other phonon modes, which display qualitatively similar behavior. 
The phonons are fit using the Voigt profile function to deconvolute the effects of the spectrometer resolution from the phonon linewidths. 
The energy and linewidth of the mode are fit to 2-phonon anharmonic decay model~\cite{klemens1966anharmonic}
\begin{eqnarray}
\omega(T) = \omega_{0} + \omega_{R}(2n_{B}(\omega_{0}/2,T) + 1) \\
\Gamma(T) = \Gamma_{0} + \Gamma_{R}(2n_{B}(\omega_{0}/2,T) + 1)
\end{eqnarray} 
where $\omega(T)$ and $\Gamma(T)$ are the energy and linewidth of the phonon at temperature $T$, $\omega_{0}$ is the bare energy, $\Gamma_{0}$ is the residual linewidth, and $\omega_{R}$ and $\Gamma_{R}$ are fitting parameters. 
The $A_{1g}^{(2)}$ mode is well described by the standard 2-phonon anharmonic decay model, with residual linewidths consistent with previous temperature dependence measurements indicating good bulk crystal quality, see Table~\ref{table:I}. 

The discontinuity of the crystal structure at the surface of \BiAll has several important consequences. 
Since inversion symmetry is broken at the surface, the distinction between Raman and IR-active phonons is lifted and both types of modes are observable using the Raman probe. 
The optical phonons at the surface are described by the irreducible representations of the $\bar{\Gamma}$ point. 
By making the approximation that the crystal structure at the surface is comparable with the bulk crystal structure, the surface phonons are classified according to the corresponding bulk phonons they branch off from. 
Under this assumption, the irreducible representations of the optical surface phonons are $4A_{1} + 4E$. 
In the notation used throughout this paper, $W^{(1,2)}$ refers to the Raman-related surface modes and $W^{(3,4)}$ refers to the IR-related surface modes, where $W = A_1$ or $E$; for example, the surface mode counterparts of the $A_{2u}^{(1)}$ and $A_{2u}^{(2)}$ modes are $A_{1}^{(3)}$ and $A_{1}^{(4)}$. 
A second important consequence is that the change in the crystal field potential at the surface will be reflected in the differences between the bulk and surface Raman spectra. 
For example, the energy and linewidth of the surface modes with shift relative to their bulk counterparts, indicating weaker or stronger bond strengths. 
Thus, information about the electronic/vibration properties at the surface may be infer from the surface Raman spectra. 

The most distinctive surface related features are the $A_1^{(3)}$ and $A_1^{(4)}$ surface phonons. 
In \BiSe and \BiTe, the $A_{1}^{(4)}$ mode is the strongest surface mode~\cite{kung2017surface,boulares2018surface} whereas in \BiTeSe the $A_{1}^{(3)}$ mode is the strongest. 
However, the majority of surface phonon modes are more than an order of magnitude weaker than their bulk counterparts, making their observation difficult. 
In order to aid their observation, we used polarization-resolved resonant Raman data to selectively probe the different irreducible representations. 
The previous Raman study used this method to identify the surface phonon modes in \BiSe~\cite{kung2017surface}, therefore, we restrict our analysis to the surface phonons in \BiTe and \BiTeSe. 

\begin{figure*}[t]%
	\centering
	\includegraphics[width=0.9\linewidth]{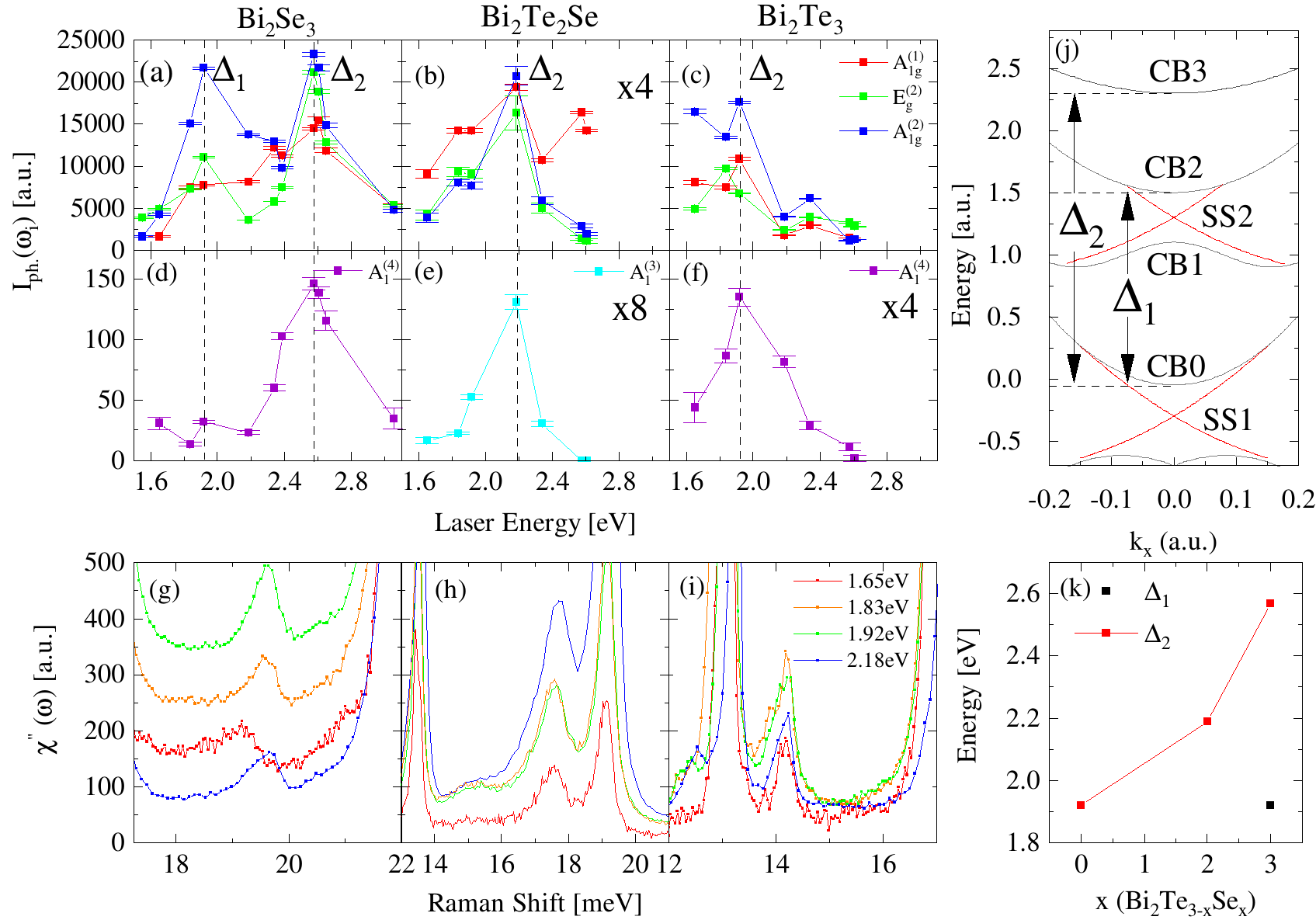}%
	\caption{Resonant Raman excitation profile (RREP) of the major low energy spectral features in \BiAll. 
		(a)-(c) Integrated intensity of the major bulk phonons, and (d)-(f) the $A_{1}^{(3)}$ (\BiSe and \BiTe) or $A_{1}^{(4)}$ (\BiTeSe) surface phonon plotted against excitation energy $\omega_{i}$. 
		The dotted vertical lines are to guide the reader's eye. 
		(g)-(i) Raman spectra for \BiAll in the spectral region of the $A_{1}^{(3)}$ and $A_{1}^{(4)}$ surface phonons with  optical corrections, see Eqs.\,(\ref{SE}) and (\ref{Corr}). 
        The PL signal was approximated to a linear background and subtracted accordingly. 
		(j) Model of the electronic band structure near the $\Gamma$ point for \BiAll. 
		The lowest energy surface states (SS1), and unoccupied topological surface states (SS2) are depicted by red lines~\cite{soifer2019bandresolved,eremeev2013new,niesner2012unoccupied,nurmamat2013unoccupied,kung2017chiral,kung2019observation}. 
		The bulk bands are shown in gray. 
		(k) Evolution of the $\Delta_{1}$ and $\Delta_{2}$ energy gaps as a function of Se concentration ($x$). 
	}
	\label{excit_profile}%
\end{figure*}

In Fig.\,\ref{surfacephonons}(a-b), we show the Raman response function of \BiTeSe in the RR and RL scattering geometries, respectively. 
The $A_1^{(1)}$ mode (7.3\,meV) is red-shifted ($\sim0.7$\,meV) from the $A_{1g}^{(1)}$ mode. 
The $A_1^{(2)}$ mode is obscured by the $A_{1g}^{*}$ feature, therefore cannot be detected by this method. 
The $E^{(1)}$ and $E^{(2)}$ modes are not sufficiently red-shifted from their bulk counterparts to be detected.
Although to the best of our knowledge there is no spectroscopic data of the $A_{2u}$ phonons in \BiTeSe, lattice dynamics calculations of \BiTeSe using density functional theory (DFT)~\cite{shi2015connecting} suggest that the $A_{1}$ features at 15.2\,meV and 16.7\,meV are the $A_1^{(3)}$ and $A_1^{(4)}$ modes, respectively. 
Similarly, the $E$ features at 7.8\,meV and 14.9\,meV are the $E^{(3)}$ and $E^{(4)}$ modes, respectively~\cite{reijnders2014optical,dipietro2012optical,aleshchenko2014infrared,akrap2012optical}.
In addition, we detected several vibrational modes with spectral intensities comparable to the surface phonons, labeled $E^{*}$, that were previously reported from IR spectroscopy measurements~\cite{akrap2012optical}, though their origin remains unclear. 

In Fig.\,\ref{surfacephonons}(c-d), we show the Raman spectra of \BiTe in the RR and RL scattering geometries, respectively. 
The $A_{1}^{(1)}$ (7.5\,meV) and $A_{1}^{(2)}$ (16.7\,meV) modes are red-shifted from their $A_{1g}$-counterparts by 0.3\,meV and 0.5\,meV, respectively. 
The $A_{1}^{(3)}$ (10.3\,meV) and $A_{1}^{(4)}$ (14.0\,meV) modes red-shifted from their IR-counterparts by 1.4\,meV and 0.9\,meV, respectively~\cite{richter1977raman}. 
Similar to the case of \BiTeSe, the $E^{(1)}$ and $E^{(2)}$ phonons are obscured by their bulk counterparts. 
The $E^{(3)}$ and $E^{(4)}$ surface modes appear at 7.1\,meV and 12.3\,meV, respectively. 

The shifts in energy of the surface phonons relative to their bulk counterparts can be understood in the following way. 
The small red-shifting of the $A_{1}$ modes is due to the lack of interlayer van der Waals restorative force on the vacuum side of the surface layer~\cite{zhang2011raman}. 
The lack of energy shifting by the $E$-modes suggests that the surface termination does not significantly alter the in-plane bond dynamics. 
We may infer, based on the similar character of the surface phonons to their bulk counter parts, that the surface lattice dynamics of \BiAll is nearly identical to the bulk. 
This is in stark contrast to conventional 3D materials where the surface termination radically alters the structural and chemical properties of the top-most atomic layers.

To further investigate the spectral characteristics of the surface and bulk phonons we collected the data for varying excitation frequency $\omega_{i}$. 
The relevant spectral characteristic of interest is the integrated intensity of a particular mode as a function of $\omega_{i}$, $I_{ph}(\omega_{i})$. 
The resonant enhancement factor of Raman active phonons is correlated to transitions between the initial and intermediate electronic states. 
Therefore, we may deduce the evolution of the electronic energy gaps in these materials by tracking $I_{ph}(\omega_{i})$. 
Moreover, the Fano anti-resonance effect on the surface phonons in  \BiSe~\cite{kung2017surface,boulares2018surface} can be studied in greater detail as both the Raman coupling to the surface phonons and to the electronic continua is enhanced when incoming photon energy $\omega_{i}$ resonates with the energies of interband transitions. 
The surface phonons were fit using the local fits of the bulk phonons in closest in proximity. 

In Fig.\,\ref{excit_profile}, we show the resonant Raman excitation profile (RREP) for the major spectral features in \BiAll and the corresponding electronic band gaps that can be deduced from the peaks in $I_{ph}(\omega_{i})$. 
In Figs.\,\ref{excit_profile}(a-c), we show the $I_{ph}(\omega_{i})$ for the $A_{1g}^{(1)}$, $A_{1g}^{(2)}$, and $E_{g}^{(2)}$ bulk phonon modes from $\omega_{i}=1.65-3.05$\,eV; the $E_{g}^{(1)}$ phonon mode is not included as its spectral intensity was insufficient for proper analysis. 
In Fig.\,\ref{excit_profile}(d-f), we show the $I_{ph}(\omega_{i})$ of either the $A_{1}^{(3)}$ or $A_{1}^{(4)}$ surface modes as the excitation profile of the other mode could not be determined. 
The dashed lines between the figures are meant to guide the reader's eye. 
In Figs.\,\ref{excit_profile}(g-i), we show Raman spectra for several excitations used in the RREP. 
The electronic band structure of \BiAll near the $\Gamma$-point is illustrated in Fig.\,\ref{excit_profile}(j)~\cite{zhang2009topological}, with the corresponding values of the electronic band gaps $\Delta_{1}$ and $\Delta_{2}$ plotted in Fig.\,\ref{excit_profile}(k). 

The RREPs of the phonon modes can be understood by considering the evolution of $\Delta_{1}$ and $\Delta_{2}$ with Se-concentration. 
The two resonances in the excitation profile in \BiSe at 1.9\,eV and 2.6\,eV correspond to the band gaps between CB0 and the third- and fourth-lowest conduction bands, CB2 and CB3, respectively~\cite{sobota2013direct,niesner2012unoccupied}. 
In contrast, \BiTeSe and \BiTe only possess a single peak in their excitation profiles at $\sim2.2$\,eV and $\sim1.9$\,eV, respectively, corresponding to the $\Delta_{2}$ band gap. 
This is because the $\Delta_{1}$ band gap is smaller than 1.6\,eV in these materials~\cite{niesner2012unoccupied}. 
These results are consistent with previous excitation dependent measurements of \BiSe and \BiTe~\cite{boulares2018surface}. 
However, previous studies only employed 3 laser energies, making it difficult to find the resonance peak and further correlate the resonance factor to the band structure~\cite{weber2000raman}. 

\begin{figure}[t]%
	\centering
	\includegraphics[width=0.9\linewidth]{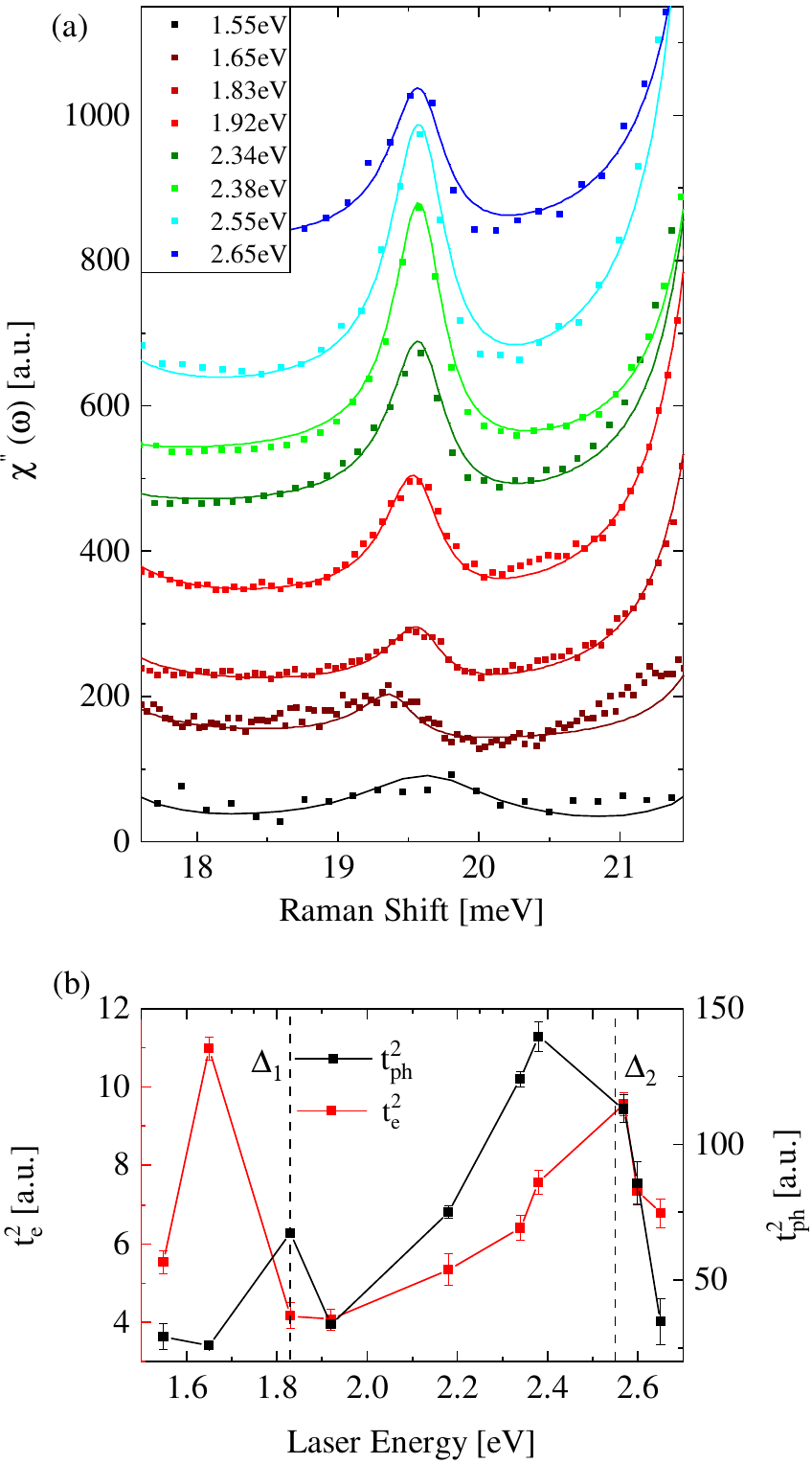}%
	\caption{\label{fano}%
		Fano resonance profile of \BiSe 
		(a) Fano resonance of the A$_{1}^{(4)}$ surface phonon in \BiSe for different $\omega_{i}$. 
		The data is shifted by a constant for each spectra for readability. 
		(b) $t_{e}^{2}$ and $t_{ph}^{2}$ as a function of excitation energy $\omega_{i}$. 
        Values are derived from the fit to the Fano lineshape. 
        The dotted vertical lines represent $\Delta_{1}$ and $\Delta_{2}$ energies, see Fig.\,\ref{excit_profile}(j). 
	}
\end{figure} 

Finally, we determined the spectral properties of the electronic continuum that induces the Fano resonance of the $A_{1}^{(4)}$ surface mode of \BiSe. 
When an electronic continuum interacts with a discrete phonon state, the Fano effect induces an asymmetrical spectral lineshape of the phonon~\cite{fano1961effects,Klein1983}. 
The degree of asymmetry depends on several factors including the electron-phonon interaction strength, $v$, the electronic density of states, $\rho(\omega)$, the phonon-light-interaction strength, $t_{ph}$, and the continuum-light-interaction strength, $t_{e}$~\cite{Klein1983,Shangfei2020}. 
Therefore, one may extract information about the continuum by tracking the phonon lineshape as a function of excitation energy $\omega_{i}$. 

In principle, a surface phonon can couple to any electronic continuum with the same symmetry where the transitioning electrons interact with the surface phonon. 
The ``electronic continuum'' in this case is the Raman continuum of of electron-hole excitations that the surface phonon energy coincides with. 
The strength of the interaction depends on several factors; in the case of \BiSe only a single continuum can couple to the $A_{1}^{(4)}$ mode. 
There are three low-energy electronic continua in \BiSe: 
the intraband surface continuum of SS1$_{-}\rightarrow$ SS1$_{+}$ transitions (where SS1$_{\pm}$ refers to the upper (lower) branches of SS1), 
the interband surface-to-bulk continua of SS1$_{-}\rightarrow$ CB0, 
and SS1$_{+}\rightarrow$ CB0 transitions continuum. 
Since the chemical potential crosses 150\,meV above the Dirac point, the onset of the intraband surface continuum is roughly 300\,meV, which is well above $A_{1}^{(4)}$ mode energy. 
The onset of the SS1$_{-}\rightarrow$ CB0 continuum will be approximately 150\,meV, with transitions between the Dirac point and the CB0 minimum~\cite{xia2009observation}. 
Thus, the $A_{1}^{(4)}$ mode falls only  within the SS1$_{+}\rightarrow$ CB0 continuum. 
Furthermore, if the SS1$_{+}\rightarrow$ CB0 continuum is selectively enhanced using resonant Raman scattering, then the effects of the other two continua are suppressed. 

The lineshape of the Raman response for a phonon mode interacting with an electronic continuum, $\chi^{''}(\omega)$, is described in the following way~\cite{Klein1983,blumberg1994investigation,kung2017surface,Shangfei2020}
\begin{eqnarray}\label{eq:reducedfano}
\chi^{''}(\omega) \sim \frac{t_{e}^{2}\pi\rho(\omega_{0}-\omega-v t_{ph}/t_{e})^{2}}{(\omega - \omega_{0})^{2} + (\gamma - v^{2}\rho\pi)^{2}}
\end{eqnarray}
where $t_{ph}$ and $t_{e}$ denotes the vertices for the light scatting processes, $v$ is the electron-phonon interaction,  $\omega_{0}$ is the bare phonon frequency, $\gamma$ is the phonon half-width at half-maximum, and $\rho$ is the density of the relevant continuum states. 
In Eq.\,\ref{eq:reducedfano}, we have assumed that $t_{e}$ and $v$ are frequency independent constants for the given continuum. 
We have also neglected the real part of the electronic self-energy and assumed that $v^{2}\rho >> \gamma$~\cite{kung2017surface}. 

To determine the spectral properties of the SS1$_{+}\rightarrow$ CB0 transition, we fit the $A_{1}^{(4)}$ mode Raman spectra taken with different $\omega_{i}$ using Eq.~\ref{eq:reducedfano}, where 
$\omega_{0}$, $\gamma$, $v$, and $\rho$ are assumed to be $\omega_{i}$-independent constants.
The factor $t_{e}$ strongly depends on $\omega_{i}$ as it undergoes enhancement via the resonant Raman process through SS1$\rightarrow$SS2 dipole transitions. 

In Fig.\,\ref{fano}(a), we display fits to the A$_{1}^{(4)}$ surface mode data in \BiSe for $\omega_{i}=1.55-2.65$\,eV excitations. 
The fits capture the strong Fano lineshape of the mode in the proximity of $\omega_{i}=1.65$\,eV excitation, the reduction of mode's asymmetry away from the resonance, and the excitation dependence of the phonon-light interaction strength. 
We obtain electron-phonon interaction strength $v=0.28$\,meV, which is similar to the value from earlier study~\cite{kung2017surface} (0.32\,meV). 

In Fig.\,\ref{fano}(b), we show the evolution of $t_{e}^{2}$ and $t_{ph}^{2}$ with excitation energy $\omega_{i}$. 
There are two resonances in the $t_{e}^{2}$ profile: one at $\omega_{i}=$1.65\,eV due to SS1-SS2 transitions, and another at $\omega_{i}=$2.5\,eV due to SS1-CB3 transitions. 
Similarly, in the $t_{ph}^{2}$  profile there is a weak resonance at 1.83\,eV and a stronger resonance at 2.55\,eV. 
The onset of the SS1-SS2 transitions is slightly below the $\Delta_{1}$ transition, which is consistent with the model of the band structure [Fig.\,\ref{excit_profile}(j)]. 
The offset between the lower energy phononic and electronic resonances is likely because the surface phonon couples stronger to SS1-CB2 transitions than to SS1-SS2 transitions. 
The pronounced Fano lineshape of the A$_{1}^{(4)}$ surface mode at $\omega_{i}=$1.65\,eV is result of the strong resonance in $t_e^2$ and the relatively weak resonance in $t_{ph}^2$ at this energy. 
Similarly, the A$_{1}^{(4)}$ surface mode develops a more Lorentzian lineshape at $\omega_{i}=$2.5\,eV due to the broader and weaker resonance in $t_{e}^2$ and the strong resonance in $t_{ph}^2$ at this energy. 

\section{Conclusions}
In conclusion, we have probed the surface phonons and the continuum of electronic transitions between surface conduction bands in \BiAll using resonant Raman spectroscopy. 
By using polarization-resolved resonant Raman spectroscopy, we observed the surface counterparts of the bulk IR active phonons and the $A_1$ Raman surface phonons in all three materials. 
The surface counterparts of the bulk Raman active $E_g$ phonons were not sufficiently red-shifted to be resolved for adequate analysis. 
We determined that the energies of all phonon modes consistently harden with increased Se concentration. 
In \BiSe, the A$_{1}^{(4)}$ surface phonon undergoes a strong Fano anti-resonance when the principal surface bands resonate with the next-lowest surface bands. 
This effect is not observed in \BiTeSe and \BiTe as the band gap in these materials is below the lowest laser energy of our setup. 
The quantitative similarities between the resonance profiles of the bulk and surface phonon modes is good evidence that the surface modes in the materials are related to the corresponding bulk phonons with renormalized energies due to their surface discontinuity. 

\section{Acknowledgements}
\begin{acknowledgments}
We thank R.\, Merlin and I.\,Boulares for discussions. 
The spectroscopic work (A.C.L., H.-H.K. and G.B.) was supported by the NSF under Grant No. DMR-2105001. 
The sample growth (X.W. and S.-W.C.) were supported by the DOE under Grant No. DE-FG02-07ER46382. 
The work at NICPB was supported by the European Research Council (ERC) under the European Union’s Horizon 2020 research and innovation programme Grant Agreement No. 885413. 
\end{acknowledgments}


\begin{thebibliography}{45}%
\makeatletter
\providecommand \@ifxundefined [1]{%
 \@ifx{#1\undefined}
}%
\providecommand \@ifnum [1]{%
 \ifnum #1\expandafter \@firstoftwo
 \else \expandafter \@secondoftwo
 \fi
}%
\providecommand \@ifx [1]{%
 \ifx #1\expandafter \@firstoftwo
 \else \expandafter \@secondoftwo
 \fi
}%
\providecommand \natexlab [1]{#1}%
\providecommand \enquote  [1]{``#1''}%
\providecommand \bibnamefont  [1]{#1}%
\providecommand \bibfnamefont [1]{#1}%
\providecommand \citenamefont [1]{#1}%
\providecommand \href@noop [0]{\@secondoftwo}%
\providecommand \href [0]{\begingroup \@sanitize@url \@href}%
\providecommand \@href[1]{\@@startlink{#1}\@@href}%
\providecommand \@@href[1]{\endgroup#1\@@endlink}%
\providecommand \@sanitize@url [0]{\catcode `\\12\catcode `\$12\catcode
  `\&12\catcode `\#12\catcode `\^12\catcode `\_12\catcode `\%12\relax}%
\providecommand \@@startlink[1]{}%
\providecommand \@@endlink[0]{}%
\providecommand \url  [0]{\begingroup\@sanitize@url \@url }%
\providecommand \@url [1]{\endgroup\@href {#1}{\urlprefix }}%
\providecommand \urlprefix  [0]{URL }%
\providecommand \Eprint [0]{\href }%
\providecommand \doibase [0]{https://doi.org/}%
\providecommand \selectlanguage [0]{\@gobble}%
\providecommand \bibinfo  [0]{\@secondoftwo}%
\providecommand \bibfield  [0]{\@secondoftwo}%
\providecommand \translation [1]{[#1]}%
\providecommand \BibitemOpen [0]{}%
\providecommand \bibitemStop [0]{}%
\providecommand \bibitemNoStop [0]{.\EOS\space}%
\providecommand \EOS [0]{\spacefactor3000\relax}%
\providecommand \BibitemShut  [1]{\csname bibitem#1\endcsname}%
\let\auto@bib@innerbib\@empty
\bibitem [{\citenamefont {Moore}(2010)}]{moore2010birth}%
  \BibitemOpen
  \bibfield  {author} {\bibinfo {author} {\bibfnamefont {J.~E.}\ \bibnamefont
  {Moore}},\ }\bibfield  {title} {\bibinfo {title} {{The birth of topological
  insulators}},\ }\href {https://www.nature.com/articles/nature08916}
  {\bibfield  {journal} {\bibinfo  {journal} {Nature}\ }\textbf {\bibinfo
  {volume} {464}},\ \bibinfo {pages} {194} (\bibinfo {year}
  {2010})}\BibitemShut {NoStop}%
\bibitem [{\citenamefont {Hasan}\ and\ \citenamefont
  {Kane}(2010)}]{hasan2010colloquium}%
  \BibitemOpen
  \bibfield  {author} {\bibinfo {author} {\bibfnamefont {M.~Z.}\ \bibnamefont
  {Hasan}}\ and\ \bibinfo {author} {\bibfnamefont {C.~L.}\ \bibnamefont
  {Kane}},\ }\bibfield  {title} {\bibinfo {title} {{Colloquium: Topological
  insulators}},\ }\href {https://doi.org/10.1103/RevModPhys.82.3045} {\bibfield
   {journal} {\bibinfo  {journal} {Rev. Mod. Phys.}\ }\textbf {\bibinfo
  {volume} {82}},\ \bibinfo {pages} {3045} (\bibinfo {year}
  {2010})}\BibitemShut {NoStop}%
\bibitem [{\citenamefont {Fu}\ \emph {et~al.}(2007)\citenamefont {Fu},
  \citenamefont {Kane},\ and\ \citenamefont {Mele}}]{fu2007topological}%
  \BibitemOpen
  \bibfield  {author} {\bibinfo {author} {\bibfnamefont {L.}~\bibnamefont
  {Fu}}, \bibinfo {author} {\bibfnamefont {C.~L.}\ \bibnamefont {Kane}},\ and\
  \bibinfo {author} {\bibfnamefont {E.~J.}\ \bibnamefont {Mele}},\ }\bibfield
  {title} {\bibinfo {title} {Topological insulators in three dimensions},\
  }\href {https://doi.org/10.1103/PhysRevLett.98.106803} {\bibfield  {journal}
  {\bibinfo  {journal} {Phys. Rev. Lett.}\ }\textbf {\bibinfo {volume} {98}},\
  \bibinfo {pages} {106803} (\bibinfo {year} {2007})}\BibitemShut {NoStop}%
\bibitem [{\citenamefont {Roushan}\ \emph {et~al.}(2009)\citenamefont
  {Roushan}, \citenamefont {Seo}, \citenamefont {Parker}, \citenamefont {Hor},
  \citenamefont {Hsieh}, \citenamefont {Qian}, \citenamefont {Richardella},
  \citenamefont {Hasan}, \citenamefont {Cava},\ and\ \citenamefont
  {Yazdani}}]{roushan2009topological}%
  \BibitemOpen
  \bibfield  {author} {\bibinfo {author} {\bibfnamefont {P.}~\bibnamefont
  {Roushan}}, \bibinfo {author} {\bibfnamefont {J.}~\bibnamefont {Seo}},
  \bibinfo {author} {\bibfnamefont {C.~V.}\ \bibnamefont {Parker}}, \bibinfo
  {author} {\bibfnamefont {Y.~S.}\ \bibnamefont {Hor}}, \bibinfo {author}
  {\bibfnamefont {D.}~\bibnamefont {Hsieh}}, \bibinfo {author} {\bibfnamefont
  {D.}~\bibnamefont {Qian}}, \bibinfo {author} {\bibfnamefont {A.}~\bibnamefont
  {Richardella}}, \bibinfo {author} {\bibfnamefont {M.~Z.}\ \bibnamefont
  {Hasan}}, \bibinfo {author} {\bibfnamefont {R.~J.}\ \bibnamefont {Cava}},\
  and\ \bibinfo {author} {\bibfnamefont {A.}~\bibnamefont {Yazdani}},\
  }\bibfield  {title} {\bibinfo {title} {{Topological surface states protected
  from backscattering by chiral spin texture}},\ }\href
  {https://www.nature.com/articles/nature08308} {\bibfield  {journal} {\bibinfo
   {journal} {Nature}\ }\textbf {\bibinfo {volume} {460}},\ \bibinfo {pages}
  {1106} (\bibinfo {year} {2009})}\BibitemShut {NoStop}%
\bibitem [{\citenamefont {\ifmmode \check{Z}\else
  \v{Z}\fi{}uti\ifmmode~\acute{c}\else \'{c}\fi{}}\ \emph
  {et~al.}(2004)\citenamefont {\ifmmode \check{Z}\else
  \v{Z}\fi{}uti\ifmmode~\acute{c}\else \'{c}\fi{}}, \citenamefont {Fabian},\
  and\ \citenamefont {Das~Sarma}}]{zuti2004spintronics}%
  \BibitemOpen
  \bibfield  {author} {\bibinfo {author} {\bibfnamefont {I.}~\bibnamefont
  {\ifmmode \check{Z}\else \v{Z}\fi{}uti\ifmmode~\acute{c}\else \'{c}\fi{}}},
  \bibinfo {author} {\bibfnamefont {J.}~\bibnamefont {Fabian}},\ and\ \bibinfo
  {author} {\bibfnamefont {S.}~\bibnamefont {Das~Sarma}},\ }\bibfield  {title}
  {\bibinfo {title} {Spintronics: Fundamentals and applications},\ }\href
  {https://doi.org/10.1103/RevModPhys.76.323} {\bibfield  {journal} {\bibinfo
  {journal} {Rev. Mod. Phys.}\ }\textbf {\bibinfo {volume} {76}},\ \bibinfo
  {pages} {323} (\bibinfo {year} {2004})}\BibitemShut {NoStop}%
\bibitem [{\citenamefont {Linder}\ \emph {et~al.}(2009)\citenamefont {Linder},
  \citenamefont {Yokoyama},\ and\ \citenamefont
  {Sudb\o{}}}]{linder2009anomalous}%
  \BibitemOpen
  \bibfield  {author} {\bibinfo {author} {\bibfnamefont {J.}~\bibnamefont
  {Linder}}, \bibinfo {author} {\bibfnamefont {T.}~\bibnamefont {Yokoyama}},\
  and\ \bibinfo {author} {\bibfnamefont {A.}~\bibnamefont {Sudb\o{}}},\
  }\bibfield  {title} {\bibinfo {title} {{Anomalous finite size effects on
  surface states in the topological insulator
  ${\text{Bi}}_{2}{\text{Se}}_{3}$}},\ }\href
  {https://doi.org/10.1103/PhysRevB.80.205401} {\bibfield  {journal} {\bibinfo
  {journal} {Phys. Rev. B}\ }\textbf {\bibinfo {volume} {80}},\ \bibinfo
  {pages} {205401} (\bibinfo {year} {2009})}\BibitemShut {NoStop}%
\bibitem [{\citenamefont {Park}\ \emph {et~al.}(2010)\citenamefont {Park},
  \citenamefont {Jung}, \citenamefont {Kim}, \citenamefont {Song},
  \citenamefont {Kim}, \citenamefont {Kimura}, \citenamefont {Lee},\ and\
  \citenamefont {Hur}}]{park2010quasiparticle}%
  \BibitemOpen
  \bibfield  {author} {\bibinfo {author} {\bibfnamefont {S.~R.}\ \bibnamefont
  {Park}}, \bibinfo {author} {\bibfnamefont {W.~S.}\ \bibnamefont {Jung}},
  \bibinfo {author} {\bibfnamefont {C.}~\bibnamefont {Kim}}, \bibinfo {author}
  {\bibfnamefont {D.~J.}\ \bibnamefont {Song}}, \bibinfo {author}
  {\bibfnamefont {C.}~\bibnamefont {Kim}}, \bibinfo {author} {\bibfnamefont
  {S.}~\bibnamefont {Kimura}}, \bibinfo {author} {\bibfnamefont {K.~D.}\
  \bibnamefont {Lee}},\ and\ \bibinfo {author} {\bibfnamefont {N.}~\bibnamefont
  {Hur}},\ }\bibfield  {title} {\bibinfo {title} {{Quasiparticle scattering and
  the protected nature of the topological states in a parent topological
  insulator ${\text{Bi}}_{2}{\text{Se}}_{3}$}},\ }\href
  {https://doi.org/10.1103/PhysRevB.81.041405} {\bibfield  {journal} {\bibinfo
  {journal} {Phys. Rev. B}\ }\textbf {\bibinfo {volume} {81}},\ \bibinfo
  {pages} {041405} (\bibinfo {year} {2010})}\BibitemShut {NoStop}%
\bibitem [{\citenamefont {Nakajima}(1963)}]{nakajima1963crystal}%
  \BibitemOpen
  \bibfield  {author} {\bibinfo {author} {\bibfnamefont {S.}~\bibnamefont
  {Nakajima}},\ }\bibfield  {title} {\bibinfo {title} {{The crystal structure
  of Bi$_2$Te$_{3- x}$Se$_x$}},\ }\href
  {https://www.sciencedirect.com/science/article/abs/pii/0022369763902075}
  {\bibfield  {journal} {\bibinfo  {journal} {Journal of Physics and Chemistry
  of Solids}\ }\textbf {\bibinfo {volume} {24}},\ \bibinfo {pages} {479}
  (\bibinfo {year} {1963})}\BibitemShut {NoStop}%
\bibitem [{\citenamefont {Ribak}\ \emph {et~al.}(2016)\citenamefont {Ribak},
  \citenamefont {Chashka}, \citenamefont {Lahoud}, \citenamefont {Naamneh},
  \citenamefont {Rinott}, \citenamefont {Ein-Eli}, \citenamefont {Plumb},
  \citenamefont {Shi}, \citenamefont {Rienks},\ and\ \citenamefont
  {Kanigel}}]{ribak2016internal}%
  \BibitemOpen
  \bibfield  {author} {\bibinfo {author} {\bibfnamefont {A.}~\bibnamefont
  {Ribak}}, \bibinfo {author} {\bibfnamefont {K.~B.}\ \bibnamefont {Chashka}},
  \bibinfo {author} {\bibfnamefont {E.}~\bibnamefont {Lahoud}}, \bibinfo
  {author} {\bibfnamefont {M.}~\bibnamefont {Naamneh}}, \bibinfo {author}
  {\bibfnamefont {S.}~\bibnamefont {Rinott}}, \bibinfo {author} {\bibfnamefont
  {Y.}~\bibnamefont {Ein-Eli}}, \bibinfo {author} {\bibfnamefont {N.~C.}\
  \bibnamefont {Plumb}}, \bibinfo {author} {\bibfnamefont {M.}~\bibnamefont
  {Shi}}, \bibinfo {author} {\bibfnamefont {E.}~\bibnamefont {Rienks}},\ and\
  \bibinfo {author} {\bibfnamefont {A.}~\bibnamefont {Kanigel}},\ }\bibfield
  {title} {\bibinfo {title} {{Internal pressure in superconducting
  Cu-intercalated ${\mathrm{Bi}}_{2}{\mathrm{Se}}_{3}$}},\ }\href
  {https://doi.org/10.1103/PhysRevB.93.064505} {\bibfield  {journal} {\bibinfo
  {journal} {Phys. Rev. B}\ }\textbf {\bibinfo {volume} {93}},\ \bibinfo
  {pages} {064505} (\bibinfo {year} {2016})}\BibitemShut {NoStop}%
\bibitem [{\citenamefont {Arakane}\ \emph {et~al.}(2012)\citenamefont
  {Arakane}, \citenamefont {Sato}, \citenamefont {Souma}, \citenamefont
  {Kosaka}, \citenamefont {Nakayama}, \citenamefont {Komatsu}, \citenamefont
  {Takahashi}, \citenamefont {Ren}, \citenamefont {Segawa},\ and\ \citenamefont
  {Ando}}]{arakane2012tunable}%
  \BibitemOpen
  \bibfield  {author} {\bibinfo {author} {\bibfnamefont {T.}~\bibnamefont
  {Arakane}}, \bibinfo {author} {\bibfnamefont {T.}~\bibnamefont {Sato}},
  \bibinfo {author} {\bibfnamefont {S.}~\bibnamefont {Souma}}, \bibinfo
  {author} {\bibfnamefont {K.}~\bibnamefont {Kosaka}}, \bibinfo {author}
  {\bibfnamefont {K.}~\bibnamefont {Nakayama}}, \bibinfo {author}
  {\bibfnamefont {M.}~\bibnamefont {Komatsu}}, \bibinfo {author} {\bibfnamefont
  {T.}~\bibnamefont {Takahashi}}, \bibinfo {author} {\bibfnamefont
  {Z.}~\bibnamefont {Ren}}, \bibinfo {author} {\bibfnamefont {K.}~\bibnamefont
  {Segawa}},\ and\ \bibinfo {author} {\bibfnamefont {Y.}~\bibnamefont {Ando}},\
  }\bibfield  {title} {\bibinfo {title} {{Tunable Dirac cone in the topological
  insulator Bi$_{2-x}$Sb$_x$Te$_{3-y}$Se$_y$}},\ }\href
  {https://www.nature.com/articles/ncomms1639} {\bibfield  {journal} {\bibinfo
  {journal} {Nature communications}\ }\textbf {\bibinfo {volume} {3}},\
  \bibinfo {pages} {636} (\bibinfo {year} {2012})}\BibitemShut {NoStop}%
\bibitem [{\citenamefont {Chen}\ \emph {et~al.}(2009)\citenamefont {Chen},
  \citenamefont {Analytis}, \citenamefont {Chu}, \citenamefont {Liu},
  \citenamefont {Mo}, \citenamefont {Qi}, \citenamefont {Zhang}, \citenamefont
  {Lu}, \citenamefont {Dai}, \citenamefont {Fang}, \citenamefont {Zhang},
  \citenamefont {Fisher}, \citenamefont {Hussain},\ and\ \citenamefont
  {Shen}}]{chen2009experimental}%
  \BibitemOpen
  \bibfield  {author} {\bibinfo {author} {\bibfnamefont {Y.~L.}\ \bibnamefont
  {Chen}}, \bibinfo {author} {\bibfnamefont {J.~G.}\ \bibnamefont {Analytis}},
  \bibinfo {author} {\bibfnamefont {J.-H.}\ \bibnamefont {Chu}}, \bibinfo
  {author} {\bibfnamefont {Z.~K.}\ \bibnamefont {Liu}}, \bibinfo {author}
  {\bibfnamefont {S.-K.}\ \bibnamefont {Mo}}, \bibinfo {author} {\bibfnamefont
  {X.~L.}\ \bibnamefont {Qi}}, \bibinfo {author} {\bibfnamefont {H.~J.}\
  \bibnamefont {Zhang}}, \bibinfo {author} {\bibfnamefont {D.~H.}\ \bibnamefont
  {Lu}}, \bibinfo {author} {\bibfnamefont {X.}~\bibnamefont {Dai}}, \bibinfo
  {author} {\bibfnamefont {Z.}~\bibnamefont {Fang}}, \bibinfo {author}
  {\bibfnamefont {S.~C.}\ \bibnamefont {Zhang}}, \bibinfo {author}
  {\bibfnamefont {I.~R.}\ \bibnamefont {Fisher}}, \bibinfo {author}
  {\bibfnamefont {Z.}~\bibnamefont {Hussain}},\ and\ \bibinfo {author}
  {\bibfnamefont {Z.-X.}\ \bibnamefont {Shen}},\ }\bibfield  {title} {\bibinfo
  {title} {{Experimental Realization of a Three-Dimensional Topological
  Insulator, Bi$_2$Te$_3$}},\ }\href {https://doi.org/10.1126/science.1173034}
  {\bibfield  {journal} {\bibinfo  {journal} {Science}\ }\textbf {\bibinfo
  {volume} {325}},\ \bibinfo {pages} {178} (\bibinfo {year}
  {2009})}\BibitemShut {NoStop}%
\bibitem [{\citenamefont {Xia}\ \emph {et~al.}(2009)\citenamefont {Xia},
  \citenamefont {Qian}, \citenamefont {Hsieh}, \citenamefont {Wray},
  \citenamefont {Pal}, \citenamefont {Lin}, \citenamefont {Bansil},
  \citenamefont {Grauer}, \citenamefont {Hor}, \citenamefont {Cava} \emph
  {et~al.}}]{xia2009observation}%
  \BibitemOpen
  \bibfield  {author} {\bibinfo {author} {\bibfnamefont {Y.}~\bibnamefont
  {Xia}}, \bibinfo {author} {\bibfnamefont {D.}~\bibnamefont {Qian}}, \bibinfo
  {author} {\bibfnamefont {D.}~\bibnamefont {Hsieh}}, \bibinfo {author}
  {\bibfnamefont {L.}~\bibnamefont {Wray}}, \bibinfo {author} {\bibfnamefont
  {A.}~\bibnamefont {Pal}}, \bibinfo {author} {\bibfnamefont {H.}~\bibnamefont
  {Lin}}, \bibinfo {author} {\bibfnamefont {A.}~\bibnamefont {Bansil}},
  \bibinfo {author} {\bibfnamefont {D.}~\bibnamefont {Grauer}}, \bibinfo
  {author} {\bibfnamefont {Y.~S.}\ \bibnamefont {Hor}}, \bibinfo {author}
  {\bibfnamefont {R.~J.}\ \bibnamefont {Cava}}, \emph {et~al.},\ }\bibfield
  {title} {\bibinfo {title} {{Observation of a large-gap topological-insulator
  class with a single Dirac cone on the surface}},\ }\href
  {https://www.nature.com/articles/nphys1274} {\bibfield  {journal} {\bibinfo
  {journal} {Nature physics}\ }\textbf {\bibinfo {volume} {5}},\ \bibinfo
  {pages} {398} (\bibinfo {year} {2009})}\BibitemShut {NoStop}%
\bibitem [{\citenamefont {Niesner}\ \emph {et~al.}(2012)\citenamefont
  {Niesner}, \citenamefont {Fauster}, \citenamefont {Eremeev}, \citenamefont
  {Menshchikova}, \citenamefont {Koroteev}, \citenamefont {Protogenov},
  \citenamefont {Chulkov}, \citenamefont {Tereshchenko}, \citenamefont {Kokh},
  \citenamefont {Alekperov}, \citenamefont {Nadjafov},\ and\ \citenamefont
  {Mamedov}}]{niesner2012unoccupied}%
  \BibitemOpen
  \bibfield  {author} {\bibinfo {author} {\bibfnamefont {D.}~\bibnamefont
  {Niesner}}, \bibinfo {author} {\bibfnamefont {T.}~\bibnamefont {Fauster}},
  \bibinfo {author} {\bibfnamefont {S.~V.}\ \bibnamefont {Eremeev}}, \bibinfo
  {author} {\bibfnamefont {T.~V.}\ \bibnamefont {Menshchikova}}, \bibinfo
  {author} {\bibfnamefont {Y.~M.}\ \bibnamefont {Koroteev}}, \bibinfo {author}
  {\bibfnamefont {A.~P.}\ \bibnamefont {Protogenov}}, \bibinfo {author}
  {\bibfnamefont {E.~V.}\ \bibnamefont {Chulkov}}, \bibinfo {author}
  {\bibfnamefont {O.~E.}\ \bibnamefont {Tereshchenko}}, \bibinfo {author}
  {\bibfnamefont {K.~A.}\ \bibnamefont {Kokh}}, \bibinfo {author}
  {\bibfnamefont {O.}~\bibnamefont {Alekperov}}, \bibinfo {author}
  {\bibfnamefont {A.}~\bibnamefont {Nadjafov}},\ and\ \bibinfo {author}
  {\bibfnamefont {N.}~\bibnamefont {Mamedov}},\ }\bibfield  {title} {\bibinfo
  {title} {{Unoccupied topological states on bismuth chalcogenides}},\ }\href
  {https://doi.org/10.1103/PhysRevB.86.205403} {\bibfield  {journal} {\bibinfo
  {journal} {Phys. Rev. B}\ }\textbf {\bibinfo {volume} {86}},\ \bibinfo
  {pages} {205403} (\bibinfo {year} {2012})}\BibitemShut {NoStop}%
\bibitem [{\citenamefont {Kung}\ \emph
  {et~al.}(2017{\natexlab{a}})\citenamefont {Kung}, \citenamefont {Salehi},
  \citenamefont {Boulares}, \citenamefont {Kemper}, \citenamefont {Koirala},
  \citenamefont {Brahlek}, \citenamefont {Lo\ifmmode \check{s}\else
  \v{s}\fi{}\ifmmode~\check{t}\else \v{t}\fi{}\'ak}, \citenamefont {Uher},
  \citenamefont {Merlin}, \citenamefont {Wang}, \citenamefont {Cheong},
  \citenamefont {Oh},\ and\ \citenamefont {Blumberg}}]{kung2017surface}%
  \BibitemOpen
  \bibfield  {author} {\bibinfo {author} {\bibfnamefont {H.-H.}\ \bibnamefont
  {Kung}}, \bibinfo {author} {\bibfnamefont {M.}~\bibnamefont {Salehi}},
  \bibinfo {author} {\bibfnamefont {I.}~\bibnamefont {Boulares}}, \bibinfo
  {author} {\bibfnamefont {A.~F.}\ \bibnamefont {Kemper}}, \bibinfo {author}
  {\bibfnamefont {N.}~\bibnamefont {Koirala}}, \bibinfo {author} {\bibfnamefont
  {M.}~\bibnamefont {Brahlek}}, \bibinfo {author} {\bibfnamefont
  {P.}~\bibnamefont {Lo\ifmmode \check{s}\else
  \v{s}\fi{}\ifmmode~\check{t}\else \v{t}\fi{}\'ak}}, \bibinfo {author}
  {\bibfnamefont {C.}~\bibnamefont {Uher}}, \bibinfo {author} {\bibfnamefont
  {R.}~\bibnamefont {Merlin}}, \bibinfo {author} {\bibfnamefont
  {X.}~\bibnamefont {Wang}}, \bibinfo {author} {\bibfnamefont {S.-W.}\
  \bibnamefont {Cheong}}, \bibinfo {author} {\bibfnamefont {S.}~\bibnamefont
  {Oh}},\ and\ \bibinfo {author} {\bibfnamefont {G.}~\bibnamefont {Blumberg}},\
  }\bibfield  {title} {\bibinfo {title} {{Surface vibrational modes of the
  topological insulator ${\mathrm{Bi}}_{2}{\mathrm{Se}}_{3}$ observed by Raman
  spectroscopy}},\ }\href {https://doi.org/10.1103/PhysRevB.95.245406}
  {\bibfield  {journal} {\bibinfo  {journal} {Phys. Rev. B}\ }\textbf {\bibinfo
  {volume} {95}},\ \bibinfo {pages} {245406} (\bibinfo {year}
  {2017}{\natexlab{a}})}\BibitemShut {NoStop}%
\bibitem [{\citenamefont {Boulares}\ \emph {et~al.}(2018)\citenamefont
  {Boulares}, \citenamefont {Shi}, \citenamefont {Kioupakis}, \citenamefont
  {Lo{\v{s}}t'{\'a}k}, \citenamefont {Uher},\ and\ \citenamefont
  {Merlin}}]{boulares2018surface}%
  \BibitemOpen
  \bibfield  {author} {\bibinfo {author} {\bibfnamefont {I.}~\bibnamefont
  {Boulares}}, \bibinfo {author} {\bibfnamefont {G.}~\bibnamefont {Shi}},
  \bibinfo {author} {\bibfnamefont {E.}~\bibnamefont {Kioupakis}}, \bibinfo
  {author} {\bibfnamefont {P.}~\bibnamefont {Lo{\v{s}}t'{\'a}k}}, \bibinfo
  {author} {\bibfnamefont {C.}~\bibnamefont {Uher}},\ and\ \bibinfo {author}
  {\bibfnamefont {R.}~\bibnamefont {Merlin}},\ }\bibfield  {title} {\bibinfo
  {title} {{Surface phonons in the topological insulators Bi$_2$Se$_3$ and
  Bi$_2$Te$_3$}},\ }\href
  {https://www.sciencedirect.com/science/article/pii/S0038109817304076}
  {\bibfield  {journal} {\bibinfo  {journal} {Solid State Communications}\
  }\textbf {\bibinfo {volume} {271}},\ \bibinfo {pages} {1} (\bibinfo {year}
  {2018})}\BibitemShut {NoStop}%
\bibitem [{\citenamefont {Dai}\ \emph {et~al.}(2016)\citenamefont {Dai},
  \citenamefont {West}, \citenamefont {Wang}, \citenamefont {Wang},
  \citenamefont {Kwok}, \citenamefont {Cheong}, \citenamefont {Zhang},\ and\
  \citenamefont {Wu}}]{dai2016toward}%
  \BibitemOpen
  \bibfield  {author} {\bibinfo {author} {\bibfnamefont {J.}~\bibnamefont
  {Dai}}, \bibinfo {author} {\bibfnamefont {D.}~\bibnamefont {West}}, \bibinfo
  {author} {\bibfnamefont {X.}~\bibnamefont {Wang}}, \bibinfo {author}
  {\bibfnamefont {Y.}~\bibnamefont {Wang}}, \bibinfo {author} {\bibfnamefont
  {D.}~\bibnamefont {Kwok}}, \bibinfo {author} {\bibfnamefont {S.-W.}\
  \bibnamefont {Cheong}}, \bibinfo {author} {\bibfnamefont {S.~B.}\
  \bibnamefont {Zhang}},\ and\ \bibinfo {author} {\bibfnamefont
  {W.}~\bibnamefont {Wu}},\ }\bibfield  {title} {\bibinfo {title} {{Toward the
  Intrinsic Limit of the Topological Insulator
  ${\mathrm{Bi}}_{2}{\mathrm{Se}}_{3}$}},\ }\href
  {https://doi.org/10.1103/PhysRevLett.117.106401} {\bibfield  {journal}
  {\bibinfo  {journal} {Phys. Rev. Lett.}\ }\textbf {\bibinfo {volume} {117}},\
  \bibinfo {pages} {106401} (\bibinfo {year} {2016})}\BibitemShut {NoStop}%
\bibitem [{\citenamefont {Wang}\ \emph {et~al.}(2013)\citenamefont {Wang},
  \citenamefont {Huang}, \citenamefont {Thimmaiah}, \citenamefont {Alam},
  \citenamefont {Bud'ko}, \citenamefont {Kaminski}, \citenamefont {Lograsso},
  \citenamefont {Canfield},\ and\ \citenamefont {Johnson}}]{wang2013native}%
  \BibitemOpen
  \bibfield  {author} {\bibinfo {author} {\bibfnamefont {L.-L.}\ \bibnamefont
  {Wang}}, \bibinfo {author} {\bibfnamefont {M.}~\bibnamefont {Huang}},
  \bibinfo {author} {\bibfnamefont {S.}~\bibnamefont {Thimmaiah}}, \bibinfo
  {author} {\bibfnamefont {A.}~\bibnamefont {Alam}}, \bibinfo {author}
  {\bibfnamefont {S.~L.}\ \bibnamefont {Bud'ko}}, \bibinfo {author}
  {\bibfnamefont {A.}~\bibnamefont {Kaminski}}, \bibinfo {author}
  {\bibfnamefont {T.~A.}\ \bibnamefont {Lograsso}}, \bibinfo {author}
  {\bibfnamefont {P.}~\bibnamefont {Canfield}},\ and\ \bibinfo {author}
  {\bibfnamefont {D.~D.}\ \bibnamefont {Johnson}},\ }\bibfield  {title}
  {\bibinfo {title} {{Native defects in tetradymite
  Bi${}_{2}$(Te${}_{x}$Se${}_{3\ensuremath{-}x}$) topological insulators}},\
  }\href {https://doi.org/10.1103/PhysRevB.87.125303} {\bibfield  {journal}
  {\bibinfo  {journal} {Phys. Rev. B}\ }\textbf {\bibinfo {volume} {87}},\
  \bibinfo {pages} {125303} (\bibinfo {year} {2013})}\BibitemShut {NoStop}%
\bibitem [{\citenamefont {Kung}\ \emph
  {et~al.}(2017{\natexlab{b}})\citenamefont {Kung}, \citenamefont {Maiti},
  \citenamefont {Wang}, \citenamefont {Cheong}, \citenamefont {Maslov},\ and\
  \citenamefont {Blumberg}}]{kung2017chiral}%
  \BibitemOpen
  \bibfield  {author} {\bibinfo {author} {\bibfnamefont {H.-H.}\ \bibnamefont
  {Kung}}, \bibinfo {author} {\bibfnamefont {S.}~\bibnamefont {Maiti}},
  \bibinfo {author} {\bibfnamefont {X.}~\bibnamefont {Wang}}, \bibinfo {author}
  {\bibfnamefont {S.-W.}\ \bibnamefont {Cheong}}, \bibinfo {author}
  {\bibfnamefont {D.~L.}\ \bibnamefont {Maslov}},\ and\ \bibinfo {author}
  {\bibfnamefont {G.}~\bibnamefont {Blumberg}},\ }\bibfield  {title} {\bibinfo
  {title} {{Chiral Spin Mode on the Surface of a Topological Insulator}},\
  }\href {https://doi.org/10.1103/PhysRevLett.119.136802} {\bibfield  {journal}
  {\bibinfo  {journal} {Phys. Rev. Lett.}\ }\textbf {\bibinfo {volume} {119}},\
  \bibinfo {pages} {136802} (\bibinfo {year} {2017}{\natexlab{b}})}\BibitemShut
  {NoStop}%
\bibitem [{\citenamefont {Kung}\ \emph {et~al.}(2019)\citenamefont {Kung},
  \citenamefont {Goyal}, \citenamefont {Maslov}, \citenamefont {Wang},
  \citenamefont {Lee}, \citenamefont {Kemper}, \citenamefont {Cheong},\ and\
  \citenamefont {Blumberg}}]{kung2019observation}%
  \BibitemOpen
  \bibfield  {author} {\bibinfo {author} {\bibfnamefont {H.-H.}\ \bibnamefont
  {Kung}}, \bibinfo {author} {\bibfnamefont {A.~P.}\ \bibnamefont {Goyal}},
  \bibinfo {author} {\bibfnamefont {D.~L.}\ \bibnamefont {Maslov}}, \bibinfo
  {author} {\bibfnamefont {X.}~\bibnamefont {Wang}}, \bibinfo {author}
  {\bibfnamefont {A.}~\bibnamefont {Lee}}, \bibinfo {author} {\bibfnamefont
  {A.~F.}\ \bibnamefont {Kemper}}, \bibinfo {author} {\bibfnamefont {S.-W.}\
  \bibnamefont {Cheong}},\ and\ \bibinfo {author} {\bibfnamefont
  {G.}~\bibnamefont {Blumberg}},\ }\bibfield  {title} {\bibinfo {title}
  {{Observation of chiral surface excitons in a topological insulator}
  {Bi$_{2}$Se$_{3}$}},\ }\href {https://doi.org/10.1073/pnas.1813514116}
  {\bibfield  {journal} {\bibinfo  {journal} {Proceedings of the National
  Academy of Sciences}\ }\textbf {\bibinfo {volume} {116}},\ \bibinfo {pages}
  {4006} (\bibinfo {year} {2019})}\BibitemShut {NoStop}%
\bibitem [{\citenamefont {McIver}\ \emph {et~al.}(2012)\citenamefont {McIver},
  \citenamefont {Hsieh}, \citenamefont {Drapcho}, \citenamefont {Torchinsky},
  \citenamefont {Gardner}, \citenamefont {Lee},\ and\ \citenamefont
  {Gedik}}]{mciver2012theoretical}%
  \BibitemOpen
  \bibfield  {author} {\bibinfo {author} {\bibfnamefont {J.~W.}\ \bibnamefont
  {McIver}}, \bibinfo {author} {\bibfnamefont {D.}~\bibnamefont {Hsieh}},
  \bibinfo {author} {\bibfnamefont {S.~G.}\ \bibnamefont {Drapcho}}, \bibinfo
  {author} {\bibfnamefont {D.~H.}\ \bibnamefont {Torchinsky}}, \bibinfo
  {author} {\bibfnamefont {D.~R.}\ \bibnamefont {Gardner}}, \bibinfo {author}
  {\bibfnamefont {Y.~S.}\ \bibnamefont {Lee}},\ and\ \bibinfo {author}
  {\bibfnamefont {N.}~\bibnamefont {Gedik}},\ }\bibfield  {title} {\bibinfo
  {title} {{Theoretical and experimental study of second harmonic generation
  from the surface of the topological insulator Bi$_{2}$Se$_{3}$}},\ }\href
  {https://doi.org/10.1103/PhysRevB.86.035327} {\bibfield  {journal} {\bibinfo
  {journal} {Phys. Rev. B}\ }\textbf {\bibinfo {volume} {86}},\ \bibinfo
  {pages} {035327} (\bibinfo {year} {2012})}\BibitemShut {NoStop}%
\bibitem [{\citenamefont {Aleshchenko}\ \emph {et~al.}(2014)\citenamefont
  {Aleshchenko}, \citenamefont {Muratov}, \citenamefont {Pavlova},
  \citenamefont {Selivanov},\ and\ \citenamefont
  {Chizhevskii}}]{aleshchenko2014infrared}%
  \BibitemOpen
  \bibfield  {author} {\bibinfo {author} {\bibfnamefont {Y.~A.}\ \bibnamefont
  {Aleshchenko}}, \bibinfo {author} {\bibfnamefont {A.~V.}\ \bibnamefont
  {Muratov}}, \bibinfo {author} {\bibfnamefont {V.~V.}\ \bibnamefont
  {Pavlova}}, \bibinfo {author} {\bibfnamefont {Y.~G.}\ \bibnamefont
  {Selivanov}},\ and\ \bibinfo {author} {\bibfnamefont {E.~G.}\ \bibnamefont
  {Chizhevskii}},\ }\bibfield  {title} {\bibinfo {title} {{Infrared
  spectroscopy of Bi$_{2}$Te$_2$Se}},\ }\href
  {https://doi.org/10.1134/S0021364014040031} {\bibfield  {journal} {\bibinfo
  {journal} {JETP Letters}\ }\textbf {\bibinfo {volume} {99}},\ \bibinfo
  {pages} {187} (\bibinfo {year} {2014})}\BibitemShut {NoStop}%
\bibitem [{\citenamefont {Cui}\ \emph {et~al.}(1999)\citenamefont {Cui},
  \citenamefont {Bhat},\ and\ \citenamefont
  {Venkatasubramanian}}]{cui1999optical}%
  \BibitemOpen
  \bibfield  {author} {\bibinfo {author} {\bibfnamefont {H.}~\bibnamefont
  {Cui}}, \bibinfo {author} {\bibfnamefont {I.}~\bibnamefont {Bhat}},\ and\
  \bibinfo {author} {\bibfnamefont {R.}~\bibnamefont {Venkatasubramanian}},\
  }\bibfield  {title} {\bibinfo {title} {{Optical constants of Bi$_{2}$Te$_{3}$
  and Sb$_{2}$Te$_{3}$ measured using spectroscopic ellipsometry}},\ }\bibfield
   {journal} {\bibinfo  {journal} {Journal of Electronic Materials}\ }\textbf
  {\bibinfo {volume} {28}},\ \href {https://doi.org/10.1007/s11664-999-0247-z}
  {10.1007/s11664-999-0247-z} (\bibinfo {year} {1999})\BibitemShut {NoStop}%
\bibitem [{\citenamefont {Klemens}(1966)}]{klemens1966anharmonic}%
  \BibitemOpen
  \bibfield  {author} {\bibinfo {author} {\bibfnamefont {P.~G.}\ \bibnamefont
  {Klemens}},\ }\bibfield  {title} {\bibinfo {title} {Anharmonic decay of
  optical phonons},\ }\href {https://doi.org/10.1103/PhysRev.148.845}
  {\bibfield  {journal} {\bibinfo  {journal} {Phys. Rev.}\ }\textbf {\bibinfo
  {volume} {148}},\ \bibinfo {pages} {845} (\bibinfo {year}
  {1966})}\BibitemShut {NoStop}%
\bibitem [{\citenamefont {Richter}\ and\ \citenamefont
  {Becker}(1977)}]{richter1977raman}%
  \BibitemOpen
  \bibfield  {author} {\bibinfo {author} {\bibfnamefont {W.}~\bibnamefont
  {Richter}}\ and\ \bibinfo {author} {\bibfnamefont {C.}~\bibnamefont
  {Becker}},\ }\bibfield  {title} {\bibinfo {title} {{A Raman and far-infrared
  investigation of phonons in the rhombohedral V2--VI3 compounds Bi$_2$Te$_3$,
  Bi$_2$Se$_3$, Sb$_2$Te$_3$ and Bi$_2$(Te$_{1- x}$Se$_x$)$_{3}$ (0$<$ x$<$
  1),(Bi$_{1- y}$Sb$_{y}$)$_2$Te$_{3}$ (0$<$ y$<$ 1)}},\ }\href
  {https://onlinelibrary.wiley.com/doi/abs/10.1002/pssb.2220840226} {\bibfield
  {journal} {\bibinfo  {journal} {physica status solidi (b)}\ }\textbf
  {\bibinfo {volume} {84}},\ \bibinfo {pages} {619} (\bibinfo {year}
  {1977})}\BibitemShut {NoStop}%
\bibitem [{\citenamefont {Wang}\ and\ \citenamefont
  {Zhang}(2012)}]{wang2012phonon}%
  \BibitemOpen
  \bibfield  {author} {\bibinfo {author} {\bibfnamefont {B.-T.}\ \bibnamefont
  {Wang}}\ and\ \bibinfo {author} {\bibfnamefont {P.}~\bibnamefont {Zhang}},\
  }\bibfield  {title} {\bibinfo {title} {{Phonon spectrum and bonding
  properties of Bi$_{2}$Se$_3$: Role of strong spin-orbit interaction}},\
  }\href {https://doi.org/10.1063/1.3689759} {\bibfield  {journal} {\bibinfo
  {journal} {Applied Physics Letters}\ }\textbf {\bibinfo {volume} {100}},\
  \bibinfo {pages} {082109} (\bibinfo {year} {2012})}\BibitemShut {NoStop}%
\bibitem [{\citenamefont {Cheng}\ and\ \citenamefont
  {Ren}(2011)}]{cheng2011phonons}%
  \BibitemOpen
  \bibfield  {author} {\bibinfo {author} {\bibfnamefont {W.}~\bibnamefont
  {Cheng}}\ and\ \bibinfo {author} {\bibfnamefont {S.-F.}\ \bibnamefont
  {Ren}},\ }\bibfield  {title} {\bibinfo {title} {{Phonons of single quintuple
  Bi${}_{2}$Te${}_{3}$ and Bi${}_{2}$Se${}_{3}$ films and bulk materials}},\
  }\href {https://doi.org/10.1103/PhysRevB.83.094301} {\bibfield  {journal}
  {\bibinfo  {journal} {Phys. Rev. B}\ }\textbf {\bibinfo {volume} {83}},\
  \bibinfo {pages} {094301} (\bibinfo {year} {2011})}\BibitemShut {NoStop}%
\bibitem [{\citenamefont {Akrap}\ \emph {et~al.}(2012)\citenamefont {Akrap},
  \citenamefont {Tran}, \citenamefont {Ubaldini}, \citenamefont {Teyssier},
  \citenamefont {Giannini}, \citenamefont {van~der Marel}, \citenamefont
  {Lerch},\ and\ \citenamefont {Homes}}]{akrap2012optical}%
  \BibitemOpen
  \bibfield  {author} {\bibinfo {author} {\bibfnamefont {A.}~\bibnamefont
  {Akrap}}, \bibinfo {author} {\bibfnamefont {M.}~\bibnamefont {Tran}},
  \bibinfo {author} {\bibfnamefont {A.}~\bibnamefont {Ubaldini}}, \bibinfo
  {author} {\bibfnamefont {J.}~\bibnamefont {Teyssier}}, \bibinfo {author}
  {\bibfnamefont {E.}~\bibnamefont {Giannini}}, \bibinfo {author}
  {\bibfnamefont {D.}~\bibnamefont {van~der Marel}}, \bibinfo {author}
  {\bibfnamefont {P.}~\bibnamefont {Lerch}},\ and\ \bibinfo {author}
  {\bibfnamefont {C.~C.}\ \bibnamefont {Homes}},\ }\bibfield  {title} {\bibinfo
  {title} {{Optical properties of Bi${}_{2}$Te${}_{2}$Se at ambient and high
  pressures}},\ }\href {https://doi.org/10.1103/PhysRevB.86.235207} {\bibfield
  {journal} {\bibinfo  {journal} {Phys. Rev. B}\ }\textbf {\bibinfo {volume}
  {86}},\ \bibinfo {pages} {235207} (\bibinfo {year} {2012})}\BibitemShut
  {NoStop}%
\bibitem [{\citenamefont {Shi}\ \emph {et~al.}(2015)\citenamefont {Shi},
  \citenamefont {Parker}, \citenamefont {Du},\ and\ \citenamefont
  {Singh}}]{shi2015connecting}%
  \BibitemOpen
  \bibfield  {author} {\bibinfo {author} {\bibfnamefont {H.}~\bibnamefont
  {Shi}}, \bibinfo {author} {\bibfnamefont {D.}~\bibnamefont {Parker}},
  \bibinfo {author} {\bibfnamefont {M.-H.}\ \bibnamefont {Du}},\ and\ \bibinfo
  {author} {\bibfnamefont {D.~J.}\ \bibnamefont {Singh}},\ }\bibfield  {title}
  {\bibinfo {title} {{Connecting Thermoelectric Performance and
  Topological-Insulator Behavior: ${\mathrm{Bi}}_{2}{\mathrm{Te}}_{3}$ and
  ${\mathrm{Bi}}_{2}{\mathrm{Te}}_{2}\mathrm{Se}$ from First Principles}},\
  }\href {https://doi.org/10.1103/PhysRevApplied.3.014004} {\bibfield
  {journal} {\bibinfo  {journal} {Phys. Rev. Applied}\ }\textbf {\bibinfo
  {volume} {3}},\ \bibinfo {pages} {014004} (\bibinfo {year}
  {2015})}\BibitemShut {NoStop}%
\bibitem [{\citenamefont {Reijnders}\ \emph {et~al.}(2014)\citenamefont
  {Reijnders}, \citenamefont {Tian}, \citenamefont {Sandilands}, \citenamefont
  {Pohl}, \citenamefont {Kivlichan}, \citenamefont {Zhao}, \citenamefont {Jia},
  \citenamefont {Charles}, \citenamefont {Cava}, \citenamefont {Alidoust},
  \citenamefont {Xu}, \citenamefont {Neupane}, \citenamefont {Hasan},
  \citenamefont {Wang}, \citenamefont {Cheong},\ and\ \citenamefont
  {Burch}}]{reijnders2014optical}%
  \BibitemOpen
  \bibfield  {author} {\bibinfo {author} {\bibfnamefont {A.~A.}\ \bibnamefont
  {Reijnders}}, \bibinfo {author} {\bibfnamefont {Y.}~\bibnamefont {Tian}},
  \bibinfo {author} {\bibfnamefont {L.~J.}\ \bibnamefont {Sandilands}},
  \bibinfo {author} {\bibfnamefont {G.}~\bibnamefont {Pohl}}, \bibinfo {author}
  {\bibfnamefont {I.~D.}\ \bibnamefont {Kivlichan}}, \bibinfo {author}
  {\bibfnamefont {S.~Y.~F.}\ \bibnamefont {Zhao}}, \bibinfo {author}
  {\bibfnamefont {S.}~\bibnamefont {Jia}}, \bibinfo {author} {\bibfnamefont
  {M.~E.}\ \bibnamefont {Charles}}, \bibinfo {author} {\bibfnamefont {R.~J.}\
  \bibnamefont {Cava}}, \bibinfo {author} {\bibfnamefont {N.}~\bibnamefont
  {Alidoust}}, \bibinfo {author} {\bibfnamefont {S.}~\bibnamefont {Xu}},
  \bibinfo {author} {\bibfnamefont {M.}~\bibnamefont {Neupane}}, \bibinfo
  {author} {\bibfnamefont {M.~Z.}\ \bibnamefont {Hasan}}, \bibinfo {author}
  {\bibfnamefont {X.}~\bibnamefont {Wang}}, \bibinfo {author} {\bibfnamefont
  {S.~W.}\ \bibnamefont {Cheong}},\ and\ \bibinfo {author} {\bibfnamefont
  {K.~S.}\ \bibnamefont {Burch}},\ }\bibfield  {title} {\bibinfo {title}
  {{Optical evidence of surface state suppression in Bi-based topological
  insulators}},\ }\href {https://doi.org/10.1103/PhysRevB.89.075138} {\bibfield
   {journal} {\bibinfo  {journal} {Phys. Rev. B}\ }\textbf {\bibinfo {volume}
  {89}},\ \bibinfo {pages} {075138} (\bibinfo {year} {2014})}\BibitemShut
  {NoStop}%
\bibitem [{\citenamefont {Di~Pietro}\ \emph {et~al.}(2012)\citenamefont
  {Di~Pietro}, \citenamefont {Vitucci}, \citenamefont {Nicoletti},
  \citenamefont {Baldassarre}, \citenamefont {Calvani}, \citenamefont {Cava},
  \citenamefont {Hor}, \citenamefont {Schade},\ and\ \citenamefont
  {Lupi}}]{dipietro2012optical}%
  \BibitemOpen
  \bibfield  {author} {\bibinfo {author} {\bibfnamefont {P.}~\bibnamefont
  {Di~Pietro}}, \bibinfo {author} {\bibfnamefont {F.~M.}\ \bibnamefont
  {Vitucci}}, \bibinfo {author} {\bibfnamefont {D.}~\bibnamefont {Nicoletti}},
  \bibinfo {author} {\bibfnamefont {L.}~\bibnamefont {Baldassarre}}, \bibinfo
  {author} {\bibfnamefont {P.}~\bibnamefont {Calvani}}, \bibinfo {author}
  {\bibfnamefont {R.}~\bibnamefont {Cava}}, \bibinfo {author} {\bibfnamefont
  {Y.~S.}\ \bibnamefont {Hor}}, \bibinfo {author} {\bibfnamefont
  {U.}~\bibnamefont {Schade}},\ and\ \bibinfo {author} {\bibfnamefont
  {S.}~\bibnamefont {Lupi}},\ }\bibfield  {title} {\bibinfo {title} {Optical
  conductivity of bismuth-based topological insulators},\ }\href
  {https://doi.org/10.1103/PhysRevB.86.045439} {\bibfield  {journal} {\bibinfo
  {journal} {Phys. Rev. B}\ }\textbf {\bibinfo {volume} {86}},\ \bibinfo
  {pages} {045439} (\bibinfo {year} {2012})}\BibitemShut {NoStop}%
\bibitem [{\citenamefont {Ren}\ \emph {et~al.}(2012)\citenamefont {Ren},
  \citenamefont {Qi}, \citenamefont {Liu}, \citenamefont {Hao}, \citenamefont
  {Huang}, \citenamefont {Zou}, \citenamefont {Yang}, \citenamefont {Li},\ and\
  \citenamefont {Zhong}}]{ren2012large}%
  \BibitemOpen
  \bibfield  {author} {\bibinfo {author} {\bibfnamefont {L.}~\bibnamefont
  {Ren}}, \bibinfo {author} {\bibfnamefont {X.}~\bibnamefont {Qi}}, \bibinfo
  {author} {\bibfnamefont {Y.}~\bibnamefont {Liu}}, \bibinfo {author}
  {\bibfnamefont {G.}~\bibnamefont {Hao}}, \bibinfo {author} {\bibfnamefont
  {Z.}~\bibnamefont {Huang}}, \bibinfo {author} {\bibfnamefont
  {X.}~\bibnamefont {Zou}}, \bibinfo {author} {\bibfnamefont {L.}~\bibnamefont
  {Yang}}, \bibinfo {author} {\bibfnamefont {J.}~\bibnamefont {Li}},\ and\
  \bibinfo {author} {\bibfnamefont {J.}~\bibnamefont {Zhong}},\ }\bibfield
  {title} {\bibinfo {title} {Large-scale production of ultrathin topological
  insulator bismuth telluride nanosheets by a hydrothermal intercalation and
  exfoliation route},\ }\href {https://doi.org/10.1039/C2JM15973B} {\bibfield
  {journal} {\bibinfo  {journal} {J. Mater. Chem.}\ }\textbf {\bibinfo {volume}
  {22}},\ \bibinfo {pages} {4921} (\bibinfo {year} {2012})}\BibitemShut
  {NoStop}%
\bibitem [{\citenamefont {Tian}\ \emph {et~al.}(2016)\citenamefont {Tian},
  \citenamefont {Osterhoudt}, \citenamefont {Jia}, \citenamefont {Cava},\ and\
  \citenamefont {Burch}}]{tian2016local}%
  \BibitemOpen
  \bibfield  {author} {\bibinfo {author} {\bibfnamefont {Y.}~\bibnamefont
  {Tian}}, \bibinfo {author} {\bibfnamefont {G.~B.}\ \bibnamefont
  {Osterhoudt}}, \bibinfo {author} {\bibfnamefont {S.}~\bibnamefont {Jia}},
  \bibinfo {author} {\bibfnamefont {R.~J.}\ \bibnamefont {Cava}},\ and\
  \bibinfo {author} {\bibfnamefont {K.~S.}\ \bibnamefont {Burch}},\ }\bibfield
  {title} {\bibinfo {title} {{Local phonon mode in thermoelectric
  Bi$_2$Te$_2$Se from charge neutral antisites}},\ }\href
  {https://doi.org/10.1063/1.4941022} {\bibfield  {journal} {\bibinfo
  {journal} {Applied Physics Letters}\ }\textbf {\bibinfo {volume} {108}},\
  \bibinfo {pages} {041911} (\bibinfo {year} {2016})}\BibitemShut {NoStop}%
\bibitem [{\citenamefont {Loudon}(1965)}]{loudon1965theory}%
  \BibitemOpen
  \bibfield  {author} {\bibinfo {author} {\bibfnamefont {R.}~\bibnamefont
  {Loudon}},\ }\bibfield  {title} {\bibinfo {title} {Theory of the resonance
  raman effect in crystals},\ }\href@noop {} {\bibfield  {journal} {\bibinfo
  {journal} {Journal de Physique}\ }\textbf {\bibinfo {volume} {26}},\ \bibinfo
  {pages} {677} (\bibinfo {year} {1965})}\BibitemShut {NoStop}%
\bibitem [{\citenamefont {Blumberg}\ \emph {et~al.}(1994)\citenamefont
  {Blumberg}, \citenamefont {Klein}, \citenamefont {B{\"o}rjesson},
  \citenamefont {Liang},\ and\ \citenamefont
  {Hardy}}]{blumberg1994investigation}%
  \BibitemOpen
  \bibfield  {author} {\bibinfo {author} {\bibfnamefont {G.}~\bibnamefont
  {Blumberg}}, \bibinfo {author} {\bibfnamefont {M.}~\bibnamefont {Klein}},
  \bibinfo {author} {\bibfnamefont {L.}~\bibnamefont {B{\"o}rjesson}}, \bibinfo
  {author} {\bibfnamefont {R.}~\bibnamefont {Liang}},\ and\ \bibinfo {author}
  {\bibfnamefont {W.}~\bibnamefont {Hardy}},\ }\bibfield  {title} {\bibinfo
  {title} {{Investigation of the temperature dependence of electron and phonon
  Raman scattering in single crystal YBa$_2$Cu$_3$O$_{6.952}$}},\ }\href@noop
  {} {\bibfield  {journal} {\bibinfo  {journal} {Journal of superconductivity}\
  }\textbf {\bibinfo {volume} {7}},\ \bibinfo {pages} {445} (\bibinfo {year}
  {1994})}\BibitemShut {NoStop}%
\bibitem [{\citenamefont {Tian}\ \emph {et~al.}(2017)\citenamefont {Tian},
  \citenamefont {Jia}, \citenamefont {Cava}, \citenamefont {Zhong},
  \citenamefont {Schneeloch}, \citenamefont {Gu},\ and\ \citenamefont
  {Burch}}]{tian2017understanding}%
  \BibitemOpen
  \bibfield  {author} {\bibinfo {author} {\bibfnamefont {Y.}~\bibnamefont
  {Tian}}, \bibinfo {author} {\bibfnamefont {S.}~\bibnamefont {Jia}}, \bibinfo
  {author} {\bibfnamefont {R.~J.}\ \bibnamefont {Cava}}, \bibinfo {author}
  {\bibfnamefont {R.}~\bibnamefont {Zhong}}, \bibinfo {author} {\bibfnamefont
  {J.}~\bibnamefont {Schneeloch}}, \bibinfo {author} {\bibfnamefont
  {G.}~\bibnamefont {Gu}},\ and\ \bibinfo {author} {\bibfnamefont {K.~S.}\
  \bibnamefont {Burch}},\ }\bibfield  {title} {\bibinfo {title} {{Understanding
  the evolution of anomalous anharmonicity in
  ${\mathrm{Bi}}_{2}{\mathrm{Te}}_{3\ensuremath{-}x}{\mathrm{Se}}_{x}$}},\
  }\href {https://doi.org/10.1103/PhysRevB.95.094104} {\bibfield  {journal}
  {\bibinfo  {journal} {Phys. Rev. B}\ }\textbf {\bibinfo {volume} {95}},\
  \bibinfo {pages} {094104} (\bibinfo {year} {2017})}\BibitemShut {NoStop}%
\bibitem [{\citenamefont {Soifer}\ \emph {et~al.}(2019)\citenamefont {Soifer},
  \citenamefont {Gauthier}, \citenamefont {Kemper}, \citenamefont {Rotundu},
  \citenamefont {Yang}, \citenamefont {Xiong}, \citenamefont {Lu},
  \citenamefont {Hashimoto}, \citenamefont {Kirchmann}, \citenamefont
  {Sobota},\ and\ \citenamefont {Shen}}]{soifer2019bandresolved}%
  \BibitemOpen
  \bibfield  {author} {\bibinfo {author} {\bibfnamefont {H.}~\bibnamefont
  {Soifer}}, \bibinfo {author} {\bibfnamefont {A.}~\bibnamefont {Gauthier}},
  \bibinfo {author} {\bibfnamefont {A.~F.}\ \bibnamefont {Kemper}}, \bibinfo
  {author} {\bibfnamefont {C.~R.}\ \bibnamefont {Rotundu}}, \bibinfo {author}
  {\bibfnamefont {S.-L.}\ \bibnamefont {Yang}}, \bibinfo {author}
  {\bibfnamefont {H.}~\bibnamefont {Xiong}}, \bibinfo {author} {\bibfnamefont
  {D.}~\bibnamefont {Lu}}, \bibinfo {author} {\bibfnamefont {M.}~\bibnamefont
  {Hashimoto}}, \bibinfo {author} {\bibfnamefont {P.~S.}\ \bibnamefont
  {Kirchmann}}, \bibinfo {author} {\bibfnamefont {J.~A.}\ \bibnamefont
  {Sobota}},\ and\ \bibinfo {author} {\bibfnamefont {Z.-X.}\ \bibnamefont
  {Shen}},\ }\bibfield  {title} {\bibinfo {title} {{Band-resolved imaging of
  photocurrent in a topological insulator}},\ }\href
  {https://doi.org/10.1103/PhysRevLett.122.167401} {\bibfield  {journal}
  {\bibinfo  {journal} {Phys. Rev. Lett.}\ }\textbf {\bibinfo {volume} {122}},\
  \bibinfo {pages} {167401} (\bibinfo {year} {2019})}\BibitemShut {NoStop}%
\bibitem [{\citenamefont {Eremeev}\ \emph {et~al.}(2013)\citenamefont
  {Eremeev}, \citenamefont {Silkin}, \citenamefont {Menshchikova},
  \citenamefont {Protogenov},\ and\ \citenamefont {Chulkov}}]{eremeev2013new}%
  \BibitemOpen
  \bibfield  {author} {\bibinfo {author} {\bibfnamefont {S.}~\bibnamefont
  {Eremeev}}, \bibinfo {author} {\bibfnamefont {I.}~\bibnamefont {Silkin}},
  \bibinfo {author} {\bibfnamefont {T.}~\bibnamefont {Menshchikova}}, \bibinfo
  {author} {\bibfnamefont {A.~P.}\ \bibnamefont {Protogenov}},\ and\ \bibinfo
  {author} {\bibfnamefont {E.~V.}\ \bibnamefont {Chulkov}},\ }\bibfield
  {title} {\bibinfo {title} {{New topological surface state in layered
  topological insulators: unoccupied Dirac cone}},\ }\href
  {https://link.springer.com/article/10.1134/S0021364012240034} {\bibfield
  {journal} {\bibinfo  {journal} {JETP letters}\ }\textbf {\bibinfo {volume}
  {96}},\ \bibinfo {pages} {780} (\bibinfo {year} {2013})}\BibitemShut
  {NoStop}%
\bibitem [{\citenamefont {Nurmamat}\ \emph {et~al.}(2013)\citenamefont
  {Nurmamat}, \citenamefont {Krasovskii}, \citenamefont {Kuroda}, \citenamefont
  {Ye}, \citenamefont {Miyamoto}, \citenamefont {Nakatake}, \citenamefont
  {Okuda}, \citenamefont {Namatame}, \citenamefont {Taniguchi}, \citenamefont
  {Chulkov}, \citenamefont {Kokh}, \citenamefont {Tereshchenko},\ and\
  \citenamefont {Kimura}}]{nurmamat2013unoccupied}%
  \BibitemOpen
  \bibfield  {author} {\bibinfo {author} {\bibfnamefont {M.}~\bibnamefont
  {Nurmamat}}, \bibinfo {author} {\bibfnamefont {E.~E.}\ \bibnamefont
  {Krasovskii}}, \bibinfo {author} {\bibfnamefont {K.}~\bibnamefont {Kuroda}},
  \bibinfo {author} {\bibfnamefont {M.}~\bibnamefont {Ye}}, \bibinfo {author}
  {\bibfnamefont {K.}~\bibnamefont {Miyamoto}}, \bibinfo {author}
  {\bibfnamefont {M.}~\bibnamefont {Nakatake}}, \bibinfo {author}
  {\bibfnamefont {T.}~\bibnamefont {Okuda}}, \bibinfo {author} {\bibfnamefont
  {H.}~\bibnamefont {Namatame}}, \bibinfo {author} {\bibfnamefont
  {M.}~\bibnamefont {Taniguchi}}, \bibinfo {author} {\bibfnamefont {E.~V.}\
  \bibnamefont {Chulkov}}, \bibinfo {author} {\bibfnamefont {K.~A.}\
  \bibnamefont {Kokh}}, \bibinfo {author} {\bibfnamefont {O.~E.}\ \bibnamefont
  {Tereshchenko}},\ and\ \bibinfo {author} {\bibfnamefont {A.}~\bibnamefont
  {Kimura}},\ }\bibfield  {title} {\bibinfo {title} {{Unoccupied topological
  surface state in Bi${}_{2}$Te${}_{2}$Se}},\ }\href
  {https://doi.org/10.1103/PhysRevB.88.081301} {\bibfield  {journal} {\bibinfo
  {journal} {Phys. Rev. B}\ }\textbf {\bibinfo {volume} {88}},\ \bibinfo
  {pages} {081301} (\bibinfo {year} {2013})}\BibitemShut {NoStop}%
\bibitem [{\citenamefont {Zhang}\ \emph {et~al.}(2011)\citenamefont {Zhang},
  \citenamefont {Peng}, \citenamefont {Soni}, \citenamefont {Zhao},
  \citenamefont {Xiong}, \citenamefont {Peng}, \citenamefont {Wang},
  \citenamefont {Dresselhaus},\ and\ \citenamefont {Xiong}}]{zhang2011raman}%
  \BibitemOpen
  \bibfield  {author} {\bibinfo {author} {\bibfnamefont {J.}~\bibnamefont
  {Zhang}}, \bibinfo {author} {\bibfnamefont {Z.}~\bibnamefont {Peng}},
  \bibinfo {author} {\bibfnamefont {A.}~\bibnamefont {Soni}}, \bibinfo {author}
  {\bibfnamefont {Y.}~\bibnamefont {Zhao}}, \bibinfo {author} {\bibfnamefont
  {Y.}~\bibnamefont {Xiong}}, \bibinfo {author} {\bibfnamefont
  {B.}~\bibnamefont {Peng}}, \bibinfo {author} {\bibfnamefont {J.}~\bibnamefont
  {Wang}}, \bibinfo {author} {\bibfnamefont {M.~S.}\ \bibnamefont
  {Dresselhaus}},\ and\ \bibinfo {author} {\bibfnamefont {Q.}~\bibnamefont
  {Xiong}},\ }\bibfield  {title} {\bibinfo {title} {{Raman spectroscopy of
  few-quintuple layer topological insulator Bi$_2$Se$_3$ nanoplatelets}},\
  }\href {https://pubs.acs.org/doi/10.1021/nl200773n} {\bibfield  {journal}
  {\bibinfo  {journal} {Nano letters}\ }\textbf {\bibinfo {volume} {11}},\
  \bibinfo {pages} {2407} (\bibinfo {year} {2011})}\BibitemShut {NoStop}%
\bibitem [{\citenamefont {Zhang}\ \emph {et~al.}(2009)\citenamefont {Zhang},
  \citenamefont {Liu}, \citenamefont {Qi}, \citenamefont {Dai}, \citenamefont
  {Fang},\ and\ \citenamefont {Zhang}}]{zhang2009topological}%
  \BibitemOpen
  \bibfield  {author} {\bibinfo {author} {\bibfnamefont {H.}~\bibnamefont
  {Zhang}}, \bibinfo {author} {\bibfnamefont {C.-X.}\ \bibnamefont {Liu}},
  \bibinfo {author} {\bibfnamefont {X.-L.}\ \bibnamefont {Qi}}, \bibinfo
  {author} {\bibfnamefont {X.}~\bibnamefont {Dai}}, \bibinfo {author}
  {\bibfnamefont {Z.}~\bibnamefont {Fang}},\ and\ \bibinfo {author}
  {\bibfnamefont {S.-C.}\ \bibnamefont {Zhang}},\ }\bibfield  {title} {\bibinfo
  {title} {{Topological insulators in Bi$_2$Se$_3$, Bi$_2$Te$_3$ and
  Sb$_2$Te$_3$ with a single Dirac cone on the surface}},\ }\href
  {https://www.nature.com/articles/nphys1270} {\bibfield  {journal} {\bibinfo
  {journal} {Nature physics}\ }\textbf {\bibinfo {volume} {5}},\ \bibinfo
  {pages} {438} (\bibinfo {year} {2009})}\BibitemShut {NoStop}%
\bibitem [{\citenamefont {Sobota}\ \emph {et~al.}(2013)\citenamefont {Sobota},
  \citenamefont {Yang}, \citenamefont {Kemper}, \citenamefont {Lee},
  \citenamefont {Schmitt}, \citenamefont {Li}, \citenamefont {Moore},
  \citenamefont {Analytis}, \citenamefont {Fisher}, \citenamefont {Kirchmann},
  \citenamefont {Devereaux},\ and\ \citenamefont {Shen}}]{sobota2013direct}%
  \BibitemOpen
  \bibfield  {author} {\bibinfo {author} {\bibfnamefont {J.~A.}\ \bibnamefont
  {Sobota}}, \bibinfo {author} {\bibfnamefont {S.-L.}\ \bibnamefont {Yang}},
  \bibinfo {author} {\bibfnamefont {A.~F.}\ \bibnamefont {Kemper}}, \bibinfo
  {author} {\bibfnamefont {J.~J.}\ \bibnamefont {Lee}}, \bibinfo {author}
  {\bibfnamefont {F.~T.}\ \bibnamefont {Schmitt}}, \bibinfo {author}
  {\bibfnamefont {W.}~\bibnamefont {Li}}, \bibinfo {author} {\bibfnamefont
  {R.~G.}\ \bibnamefont {Moore}}, \bibinfo {author} {\bibfnamefont {J.~G.}\
  \bibnamefont {Analytis}}, \bibinfo {author} {\bibfnamefont {I.~R.}\
  \bibnamefont {Fisher}}, \bibinfo {author} {\bibfnamefont {P.~S.}\
  \bibnamefont {Kirchmann}}, \bibinfo {author} {\bibfnamefont {T.~P.}\
  \bibnamefont {Devereaux}},\ and\ \bibinfo {author} {\bibfnamefont {Z.-X.}\
  \bibnamefont {Shen}},\ }\bibfield  {title} {\bibinfo {title} {{Direct Optical
  Coupling to an Unoccupied Dirac Surface State in the Topological Insulator
  ${\mathrm{Bi}}_{2}{\mathrm{Se}}_{3}$}},\ }\href
  {https://doi.org/10.1103/PhysRevLett.111.136802} {\bibfield  {journal}
  {\bibinfo  {journal} {Phys. Rev. Lett.}\ }\textbf {\bibinfo {volume} {111}},\
  \bibinfo {pages} {136802} (\bibinfo {year} {2013})}\BibitemShut {NoStop}%
\bibitem [{\citenamefont {Weber}\ and\ \citenamefont
  {Merlin}(2000)}]{weber2000raman}%
  \BibitemOpen
  \bibfield  {author} {\bibinfo {author} {\bibfnamefont {W.~H.}\ \bibnamefont
  {Weber}}\ and\ \bibinfo {author} {\bibfnamefont {R.}~\bibnamefont {Merlin}},\
  }\href {https://link.springer.com/book/10.1007/978-3-662-04221-2} {\emph
  {\bibinfo {title} {Raman scattering in materials science}}},\ Vol.~\bibinfo
  {volume} {42}\ (\bibinfo  {publisher} {Springer Science \& Business Media},\
  \bibinfo {year} {2000})\ Chap.\ \bibinfo {chapter} {1.2}\BibitemShut
  {NoStop}%
\bibitem [{\citenamefont {Fano}(1961)}]{fano1961effects}%
  \BibitemOpen
  \bibfield  {author} {\bibinfo {author} {\bibfnamefont {U.}~\bibnamefont
  {Fano}},\ }\bibfield  {title} {\bibinfo {title} {Effects of configuration
  interaction on intensities and phase shifts},\ }\href
  {https://doi.org/10.1103/PhysRev.124.1866} {\bibfield  {journal} {\bibinfo
  {journal} {Phys. Rev.}\ }\textbf {\bibinfo {volume} {124}},\ \bibinfo {pages}
  {1866} (\bibinfo {year} {1961})}\BibitemShut {NoStop}%
\bibitem [{\citenamefont {edited~by Cardona}()}]{Klein1983}%
  \BibitemOpen
  \bibfield  {author} {\bibinfo {author} {\bibfnamefont {M.}~\bibnamefont
  {edited~by Cardona}},\ }\href@noop {} {\emph {\bibinfo {title} {{Electronic
  Raman scattering, in Light Scattering in Solids I}}}}\ (\bibinfo  {publisher}
  {Springer-Verlag, Berlin, 1983})\ pp.\ \bibinfo {pages}
  {147--204}\BibitemShut {NoStop}%
\bibitem [{\citenamefont {Wu}\ \emph {et~al.}(2020)\citenamefont {Wu},
  \citenamefont {Zhang}, \citenamefont {Li}, \citenamefont {Cao}, \citenamefont
  {Kung}, \citenamefont {Sefat}, \citenamefont {Ding}, \citenamefont
  {Richard},\ and\ \citenamefont {Blumberg}}]{Shangfei2020}%
  \BibitemOpen
  \bibfield  {author} {\bibinfo {author} {\bibfnamefont {S.-F.}\ \bibnamefont
  {Wu}}, \bibinfo {author} {\bibfnamefont {W.-L.}\ \bibnamefont {Zhang}},
  \bibinfo {author} {\bibfnamefont {L.}~\bibnamefont {Li}}, \bibinfo {author}
  {\bibfnamefont {H.-B.}\ \bibnamefont {Cao}}, \bibinfo {author} {\bibfnamefont
  {H.-H.}\ \bibnamefont {Kung}}, \bibinfo {author} {\bibfnamefont {A.~S.}\
  \bibnamefont {Sefat}}, \bibinfo {author} {\bibfnamefont {H.}~\bibnamefont
  {Ding}}, \bibinfo {author} {\bibfnamefont {P.}~\bibnamefont {Richard}},\ and\
  \bibinfo {author} {\bibfnamefont {G.}~\bibnamefont {Blumberg}},\ }\bibfield
  {title} {\bibinfo {title} {{Coupling of fully symmetric As phonon to
  magnetism in
  $\mathrm{Ba}{({\mathrm{Fe}}_{1\ensuremath{-}x}{\mathrm{Au}}_{x})}_{2}{\mathrm{As}}_{2}$}},\
  }\href {https://doi.org/10.1103/PhysRevB.102.014501} {\bibfield  {journal}
  {\bibinfo  {journal} {Phys. Rev. B}\ }\textbf {\bibinfo {volume} {102}},\
  \bibinfo {pages} {014501} (\bibinfo {year} {2020})}\BibitemShut {NoStop}%
\end{thebibliography}
%

\end{document}